\documentclass[twocolumn,numberedappendix,iop]{openjournal}
\usepackage{graphicx,amsmath,amssymb,amstext}
\usepackage{amsbsy,amsfonts,amsthm,color}
\usepackage[colorlinks,linkcolor=blue,citecolor=blue,urlcolor=blue ]{hyperref}
\usepackage[utf8]{inputenc}
\usepackage{float}
\usepackage{xcolor}
\usepackage{tabularx}
\usepackage{multirow}
\usepackage{ulem}
\usepackage[T1]{fontenc}
\usepackage[title]{appendix}

\newcommand\vo{\boldsymbol{\Omega}}
\newcommand\vl{\boldsymbol{\ell}}
\newcommand\vt{\boldsymbol{\theta}}
\newcommand\vk{\boldsymbol{k}}
\newcommand\vR{\boldsymbol{R}}
\newcommand\vs{\boldsymbol{r}}
\newcommand\into{\int \frac{d^2\vo}{\Omega_s}}
\newcommand\intvl{\int \frac{d^2\vl}{(2\pi)^2}}
\newcommand\intl{\int \frac{d\ell \, \ell}{2\pi}}
\newcommand\intlp{\int \frac{d\ell' \, \ell'}{2\pi}}
\newcommand\intk{\int \frac{d^3\vk}{(2\pi)^3}}
\newcommand\intkp{\int \frac{d^3\vk'}{(2\pi)^3}}
\newcommand\intvkperp{\int \frac{d^2\vk_\perp}{(2\pi)^2}}
\newcommand\intkperp{\int \frac{dk_\perp \, k_\perp}{2\pi}}
\newcommand\intkpar{\int_{-\infty}^\infty \frac{dk_\parallel}{2\pi}}

\begin{document}

\title{The DESI-Lensing Mock Challenge: \\ large-scale cosmological analysis of $3 \times 2$-pt statistics\vspace{-4em}}

\author{C.~Blake,$^{1,*}$
C.~Garcia-Quintero,$^{2,3,\dagger}$
S.~Ahlen,$^{4}$
D.~Bianchi,$^{5}$
D.~Brooks,$^{6}$
T.~Claybaugh,$^{7}$
A.~de la Macorra,$^{8}$
J.~DeRose,$^{7}$
A.~Dey,$^{9}$
P.~Doel,$^{6}$
N.~Emas,$^{1}$
S.~Ferraro,$^{7,10}$
J.~E.~Forero-Romero,$^{11,12}$
G.~Gutierrez,$^{13}$
S.~Heydenreich,$^{14}$
K.~Honscheid,$^{15,16,17}$
C.~Howlett,$^{18}$
M.~Ishak,$^{3}$
S.~Joudaki,$^{19}$
E.~Jullo,$^{20}$
R.~Kehoe,$^{21}$
D.~Kirkby,$^{22}$
A.~Kremin,$^{7}$
A.~Krolewski,$^{23,24,25}$
M.~Landriau,$^{7}$
J.~U.~Lange,$^{26,27}$
A.~Leauthaud,$^{14,28}$
M.~E.~Levi,$^{7}$
M.~Manera,$^{29,30}$
R.~Miquel,$^{31,30}$
J.~Moustakas,$^{32}$
G.~Niz,$^{33,34}$
W.~J.~Percival,$^{23,24,25}$
I.~P\'erez-R\`afols,$^{35}$
A.~Porredon,$^{36,37,17}$
G.~Rossi,$^{38}$
R.~Ruggeri,$^{1,18}$
E.~Sanchez,$^{36}$
C.~Saulder,$^{39}$
D.~Schlegel,$^{7}$
D.~Sprayberry,$^{9}$
Z.~Sun,$^{40}$
G.~Tarl\'{e},$^{41}$
and B.~A.~Weaver$^{9}$ \\
{\it (Affiliations can be found after the references)}
}
\thanks{$^*$E-mail: cblake@swin.edu.au}
\thanks{$^\dagger$NASA Einstein Fellow}

\begin{abstract}
The current generation of large galaxy surveys will test the cosmological model by combining multiple types of observational probes.  Realising the statistical promise of these new datasets requires rigorous attention to all aspects of analysis including cosmological measurements, modelling, covariance and parameter likelihood.  In this paper we present the results of an end-to-end simulation study designed to test the analysis pipeline for the combination of the Dark Energy Spectroscopic Instrument (DESI) Year 1 galaxy redshift dataset and separate weak gravitational lensing information from the Kilo-Degree Survey, Dark Energy Survey and Hyper-Suprime-Cam Survey.  Our analysis employs the $3 \times 2$-pt correlation functions including cosmic shear $\xi_\pm(\theta)$ and galaxy-galaxy lensing $\gamma_t(\theta)$, together with the projected correlation function of the spectroscopic DESI lenses, $w_p(R)$.  We build realistic simulations of these datasets including galaxy halo occupation distributions, photometric redshift errors, weights, multiplicative shear calibration biases and magnification.  We calculate the analytical covariance of these correlation functions including the Gaussian, noise and super-sample contributions, and show that our covariance determination agrees with estimates based on the ensemble of simulations.  We use a Bayesian inference platform to demonstrate that we can recover the fiducial cosmological parameters of the simulation within the statistical error margin of the experiment, investigating the sensitivity to scale cuts.  This study is the first in a sequence of papers in which we present and validate the large-scale $3 \times 2$-pt cosmological analysis of DESI-Y1.
  \\[1em]
\end{abstract}

\maketitle

\section{Introduction}

The joint analysis of multiple cosmological observables is a key approach for defining and ultimately resolving the outstanding questions of cosmology \citep[for recent reviews see, e.g.,][]{2013PhR...530...87W, 2022LRR....25....6M, 2023A&ARv..31....2H}.  These puzzles include the approximately 5$\sigma$ statistical ``tension'' evident in independent determinations of the present-day expansion rate of the Universe \citep[e.g.,][]{2016ApJ...826...56R, 2021ApJ...919...16F, 2021APh...13102605D, 2022JHEAp..34...49A} and, to a lesser extent, the 2-3$\sigma$ tension in its matter fluctuation spectrum \citep[e.g.,][]{2017MNRAS.467.3024L, 2018MNRAS.474.4894J, 2021A&A...646A.140H}.  The current generation of cosmological surveys, created by facilities such as the {\it Euclid} satellite \citep{2024arXiv240513491E}, the Vera Rubin Observatory \citep{2019ApJ...873..111I}, the Roman Space Telescope \citep{2021MNRAS.507.1746E} and the Dark Energy Spectroscopic Instrument \citep[DESI,][]{2016arXiv161100036D}, promise a further revolution in statistical precision.  However, realising the statistical promise of such cosmological tests places stringent requirements on all parts of the analysis pipeline including the data processing, error budget, cosmological models and likelihood determinations, motivating stringent tests of all these elements.

In this paper we use new, realistic cosmological simulations of combined-probe experiments, involving the galaxy large-scale structure of the Universe and weak gravitational lensing, to perform an end-to-end validation of the recovery of cosmological parameters.  Our study takes place in the context of the upcoming release of the DESI Year 1 (Y1) dataset, and its joint analysis with weak gravitational lensing catalogues provided by the Kilo-Degree Survey \citep[KiDS-1000,][]{2021A&A...645A.105G}, the Dark Energy Survey \citep[DES-Y3,][]{2021MNRAS.504.4312G} and Hyper-Suprime-Cam survey \citep[HSC-Y1,][]{2018PASJ...70S..25M}\footnote{After we completed the analysis presented in this paper, the HSC-Y3 dataset \citep{2022PASJ...74..421L} became publicly available.}.  The joint observables that may be defined from these datasets, involving auto- and cross-correlations between galaxy overdensity and cosmic shear, form a set of statistics known as the $3 \times 2$-pt correlation functions, and have formed the basis of the leading cosmological analyses presented by these lensing collaborations \citep[e.g.,][]{2018MNRAS.474.4894J, 2018PhRvD..98d3526A, 2021A&A...646A.140H, 2022PhRvD.105b3520A, 2023PhRvD.108l3521S, 2023PhRvD.108l3517M}.  These capabilities will be further enhanced by the addition of the DESI-Y1 spectroscopic dataset, which enjoys significant overlap with all three of these weak lensing surveys.  In this study we will utilise the projected clustering of the spectroscopic galaxy data as a component of the $3 \times 2$-pt correlation functions; future collaboration analyses will consider the fully 3-dimensional information including redshift-space distortions, as quantified by the clustering multipoles.

In general terms, three different approaches can be used to test and validate $3 \times 2$-pt cosmology analyses.  First, large-scale cosmological simulations may be designed to test cosmological parameter recovery, modelling accuracy or data covariance \citep[e.g.,][]{2015MNRAS.447.1319F, 2017ApJ...850...24T, 2018MNRAS.481.1337H, 2020PhRvD.102l3522P, 2020PhRvD.102h3520S, 2021A&A...646A.129J, 2022PhRvD.105l3520D}.  This is the focus of the current study.  This approach has the merit of allowing end-to-end tests with realistic complexity.  However, simulations may not fully capture all cosmological, astrophysical or observational effects that are present in the data, potentially limiting the robustness of conclusions.  Second, ``analytical'' techniques may be applied in which model variations associated with (for example) galaxy bias, matter power spectrum modelling and intrinsic alignments are propagated to variations in cosmological parameters, and analysis scale cuts are designed to minimize the impact of these modelling assumptions \citep[e.g.,][]{2017arXiv170609359K, 2021arXiv210513548K, 2022PhRvD.105b3514A, 2022PhRvD.105b3515S, 2023OJAp....6E..36D}.  The forthcoming study by \cite{NimasDESI} will present tests of this nature in the context of our DESI-Y1 analysis.  Finally, analysis variations may be considered in the parameter fitting of the real dataset.  However, the data vector is typically ``blinded'' during the pipeline development to lessen the possibility of ``confirmation bias'' \citep{2017MNRAS.465.1454H, 2020JCAP...09..052B, 2020MNRAS.494.4454M}.  This may affect the application of some tests, and changes to the analysis framework are generally limited after unblinding.  Our final DESI-Y1 $3 \times 2$-pt cosmology analysis will be presented in the forthcoming paper by \cite{AnnaDESI}.

Our current study is enabled by new large-volume cosmological $N$-body simulations, which simultaneously model the shear field due to weak gravitational lensing and the galaxy distribution evolving within the large-scale structure.  Such simulations must both cover large sky areas, to enable accurate tests of the data covariance and parameter recovery, and achieve sufficient resolution to match the lens and source densities and map important astrophysical effects such as magnification, photometric redshift error and intrinsic galaxy alignments.  We base our study on the Buzzard simulation suite \citep{2019arXiv190102401D} which we have populated with samples of DESI galaxies and KiDS-1000, DES-Y3 and HSC-Y1 sources in a manner matching the statistical properties of these datasets such as the redshift distributions, clustering of the lenses, and weights, shape noise, photometric redshift error and shear calibration bias of the sources \citep{2024OJAp....7E..57L}.  We do not include intrinsic alignments in the mock source catalogues used in this study, although we plan to do so in future.  The Buzzard simulations have also been used to validate methodologies for the DES-Y3 cosmological analysis \citep{2022PhRvD.105l3520D}.

In this study we use our survey simulations to test two main aspects of the DESI-Y1 $3 \times 2$-pt analysis pipeline.  First, we test the analytical model for the data covariance, which describes the correlated measurement errors between the $3 \times 2$-pt correlations as a function of scale and redshift.  We construct our model from the theoretical formulations of \cite{2017MNRAS.470.2100K} and \cite{2021A&A...646A.129J}, which we extend by presenting the covariances between the projected galaxy correlation function and the angular statistics.  This extension is motivated by the availability of spectroscopic redshifts for every  ``lens'' galaxy, in terms of which the projected correlation function may be measured with higher signal-to-noise than purely angular galaxy correlations.  Secondly we use these covariances, together with correlation function measurements and theoretical models within the {\tt CosmoMC} platform \citep{2013PhRvD..87j3529L}, to test the recovery of the fiducial cosmological parameters in a Bayesian likelihood framework.

Our paper is structured as follows.  In Sec.~\ref{sec:simulations} we summarise the survey simulations and correlation function measurements which form the basis of our analysis.  In Sec.~\ref{sec:covariance} we outline our calculation of the analytical data covariance of these correlation functions (which is fully detailed in Appendices \ref{sec:gaussian}, \ref{sec:supersample}, \ref{sec:noisecorr} and \ref{sec:covwp}).  In Sec.~\ref{sec:modelfits} we describe the theoretical modelling and Bayesian inference platform we use to create validation tests for the recovery of fiducial cosmological parameters, and present the results of these analyses in Sec.~\ref{sec:cosmo}.  In Sec.~\ref{sec:conc} we discuss remaining challenges and look forward to the upcoming application of our methodology to DESI-Y1 data.

\section{Survey simulations}
\label{sec:simulations}

In this section we provide an overview of the DESI-Lensing mock catalogues that we use in this paper to test the analytical covariance and cosmological model fits.  These catalogues are drawn from the Buzzard $N$-body simulations \citep{2019arXiv190102401D} and have also been utilised in the DESI-Lensing context to quantify astrophysical systematics in galaxy-galaxy lensing analyses by \cite{2024OJAp....7E..57L}, and the covariance of projected statistics by \cite{2024MNRAS.533..589Y}.  The fiducial cosmology of the Buzzard simulation is a flat $\Lambda$CDM model with $\Omega_m = 0.286$, $\Omega_b = 0.046$, $h = 0.7$, $\sigma_8 = 0.82$ and $n_s = 0.96$.  We used 8 independent simulations, each covering a quadrant of sky ($\approx 10{,}000$ deg$^2$).

\subsection{Buzzard mocks}
\label{sec:buzzard}

We select our mock catalogues from the complete galaxy population which has been constructed from the Buzzard simulations using the {\tt ADDGALS} algorithm \citep{2022ApJ...931..145W}.  We draw sub-samples from the galaxy population resembling the different DESI target classes -- the Bright Galaxy Survey \citep[BGS,][]{2023AJ....165..253H}, Luminous Red Galaxy \citep[LRG,][]{2023AJ....165...58Z} and Emission Line Galaxy \citep[ELG,][]{2023AJ....165..126R} samples -- together with weak lensing source populations resembling the Kilo-Degree Survey \citep[KiDS-1000,][]{2021A&A...645A.105G, 2021A&A...647A.124H}, Dark Energy Survey \citep[DES-Y3,][]{2021MNRAS.504.4312G, 2021MNRAS.505.4249M} and Hyper-Suprime-Cam \citep[HSC-Y1,][]{2018PASJ...70S..25M} datasets.

As described by \cite{2024OJAp....7E..57L}, the mock weak lensing surveys include source photo-$z$ errors, weights and multiplicative shear calibration biases matching the frameworks used by each weak lensing collaboration, such that the source redshift distributions, effective number density and shape noise within each tomographic bin match each weak lensing dataset.  In particular, source weights are assigned by sampling from the closest neighbours to each simulated source in the real data catalogues based on a {\tt KDTree} match in apparent magnitude space using all bands available for each survey.

\begin{figure*}
\centering
\includegraphics[width=2.0\columnwidth]{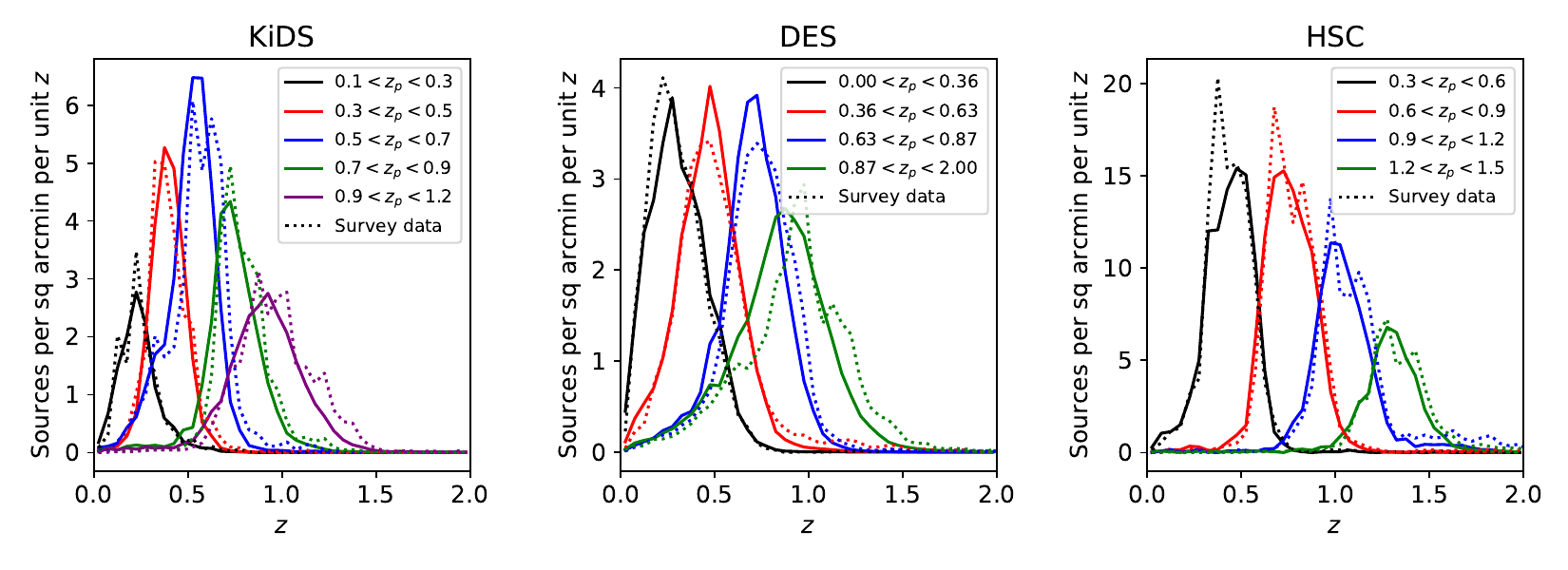}
\caption{The weighted angular density per unit redshift of the mock Buzzard source samples used in our analysis, spanning different tomographic samples for the KiDS, DES and HSC datasets as divided by photometric redshift $z_p$, and displayed as the solid lines.  These densities are compared with those estimated for the survey datasets themselves using redshift calibration samples, and shown as the dotted lines.  The ``spikes'' in the redshift distributions of the real survey datasets are caused by sample variance, given that the calibration samples only span small areas of sky.}
\label{fig:nz_source}
\end{figure*}

The mock DESI samples were populated by halo occupation distribution recipes to match the clustering and number density of the DESI Early Data Release dataset \citep{2024AJ....168...58D}, which represents a smaller but more spectroscopically complete sub-sample of the DESI-Y1 observations.  In our simulation study we do not consider the effect of spectroscopic incompleteness or fibre collisions within the DESI-Y1 sample, noting that these may be readily corrected by using appropriate weights for galaxy-galaxy lensing studies \citep{2024OJAp....7E..57L} and clustering measurements \citep{2018MNRAS.481.2338B}.  Both source and lens samples include the effects of magnification, although we do not include the effects of intrinsic alignments in this study.  Therefore the mocks do not include all effects present in the real data analysis, but nonetheless allow us to create realistic end-to-end pipeline tests.

For each source survey configuration, we divided the source and lens samples within each region into tomographic bins.  For the source surveys, we matched the tomography chosen by each weak lensing collaboration, utilising $N_{\rm tom} = (5,4,4)$ samples for (KiDS, DES, HSC), where the effective source density, shape noise and redshift distribution of each sample match that of the corresponding real lensing dataset.  The redshift distributions of the mock and survey lensing datasets are compared in Fig.~\ref{fig:nz_source}, indicating a good match.

\begin{figure}
\centering
\includegraphics[width=\columnwidth]{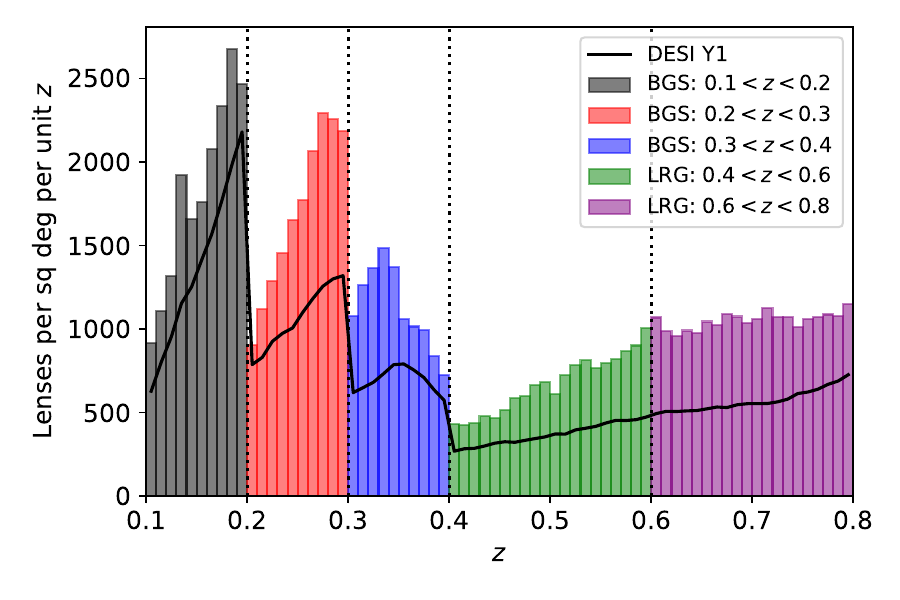}
\caption{The angular density per unit redshift of the mock Buzzard DESI  samples used in our analysis, spanning three BGS and two LRG tomographic bins, displayed as the histograms.  These densities are compared with the corresponding DESI Y1 sample, shown as the solid line.  For the purposes of this study, the mocks do not contain incompleteness effects which are present in the Y1 sample, hence have a higher number density than the data.}
\label{fig:nz_lens}
\end{figure}

For the mock DESI galaxy samples, we used $N_{\rm lens} = 5$ redshift bins with divisions $z = (0.1, 0.2, 0.3, 0.4, 0.6, 0.8)$, selecting mock BGS tracers in the redshift range $0.1 < z < 0.4$, and mock LRGs in the range $0.4 < z < 0.8$.  The angular densities per unit redshift of the mock and DESI Y1 datasets are compared in Fig.~\ref{fig:nz_lens}, where the lower number density of the data indicates incompleteness effects which are not present in the mock.  We do not currently use ELG tracers in our lensing analyses, since these are located at higher redshifts and do not serve as efficient lenses for current source samples.  However, ongoing studies are using the DESI ELG sample to test the source redshift distributions using clustering redshifts, and to place limits on intrinsic alignment models and other effects.

The BGS selection additionally includes an absolute magnitude cut $M_r < (-19.5, -20.5, -21.0)$ in the three redshift bins, creating a more homogeneous sample of lenses within each redshift interval \citep{2024MNRAS.533..589Y}.  These choices match our galaxy selection used in the lensing analysis of the DESI-Y1 dataset.  For the covariance evaluation described below, we adopt fiducial linear bias factors $b = (1.35, 1.51, 1.65, 2.21, 2.44)$ for the five lens redshift bins, based on fits to the projected correlation functions.

We note that the $0.3 < z < 0.4$ lens bin includes the transition at $z = 0.34$ between the two light cones which are used to construct the Buzzard simulations \citep{2019arXiv190102401D}.  The resulting cross-correlation coefficient between matter and galaxies is less than $1.0$ on linear scales in this region \citep{2024arXiv240704795C}, which may impact the model fits in this bin.

\subsection{Region geometry}

\begin{figure*}
\centering
\includegraphics[width=1.5\columnwidth]{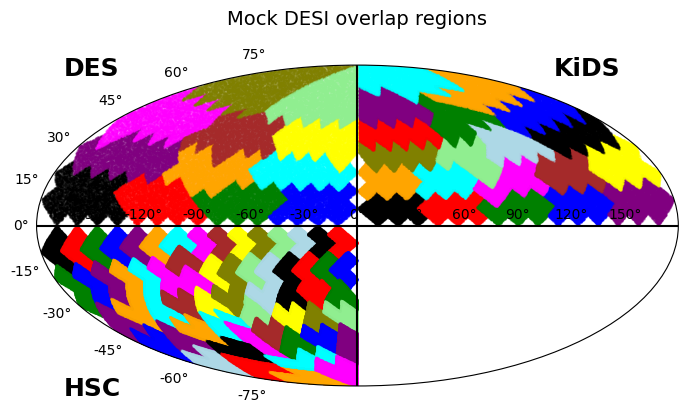}
\caption{The division of each Buzzard simulation quadrant into multiple analysis regions corresponding to the representative overlap area of DESI-Y1 and KiDS-1000, DES-Y3 and HSC-Y1.  Each quadrant is divided into 20 KiDS regions, 12 DES regions and 60 HSC regions, corresponding to 9, 15 and 3 pixels of a {\tt Healpix} $n_{\rm side} = 8$ pixelisation, respectively.  The different segmentations used to represent each weak lensing survey are displayed in separate quadrants of the figure, although they are applied to the same quadrant of Buzzard data when creating the mocks.}
\label{fig:buzzard_regions}
\end{figure*}

For the purposes of this study we divided each simulation quadrant into closely-packed regions approximating the separate overlap areas of DESI-Y1 and KiDS-1000, DES-Y3 and HSC-Y1, and we treated these regions as independent when using them to numerically estimate the covariance from correlation function measurements.  Given that the Buzzard quadrants are generated in a {\tt Healpix} $n_{\rm side} = 8$ pixelisation \citep{2019arXiv190102401D}, it is convenient to combine these pixels when constructing the regions.  Fig.~\ref{fig:buzzard_regions} displays how these pixels are grouped together to form 20 KiDS regions, 12 DES regions and 60 HSC regions from each quadrant, where these regions are comprised of 9, 15 and 3 pixels, respectively, containing area $483$, $806$ and $161$ deg$^2$ (noting again that these represent overlap areas of each lensing survey with DESI-Y1).  Whilst not exactly matching the angular footprints of these samples\footnote{We refer the reader to \cite{2024MNRAS.533..589Y} for tests of the effect of the DESI survey footprint on galaxy-galaxy lensing measurements.}, these choices allow the quadrant to be fully tiled, maximising the number of realisations, and enabling a representative test of the analytical covariance.  The 8 Buzzard mocks yielded a total of 160, 96 and 480 regions for the KiDS, DES and HSC configurations.  Although this number of regions is not sufficient to accurately determine and invert a full numerical covariance matrix, it allows us to reliably compare the diagonal elements and the off-diagonal structure with highest amplitude.  For our tests of cosmological parameter recovery, we combined these regions in different combinations, corresponding to the overlapping and non-overlapping area of these different surveys.

\subsection{Correlation function measurements}
\label{sec:corr}

We measured the components of the 3$\times$2-pt correlation functions for the different combinations of source and lens samples: the cosmic shear, $\xi_\pm(\theta)$, the average tangential shear around the lens galaxies, $\gamma_t(\theta)$, and the projected lens galaxy correlation function, $w_p(R)$.  We used the projected correlation function of DESI galaxies, in preference to the angular correlation function $w(\theta)$, to increase the clustering signal-to-noise given that every lens galaxy has a spectroscopic redshift.  This radial clustering information is lost in an angular correlation measurement owing to the projection over the full radial depth of the sample, whereas the projected correlation function is based on the 3D spatial correlation and only loses information as projected over the maximum radial separation $\Pi_\textrm{max}$ defined below.  Future work will extend these correlation statistics to consider the fully 3-dimensional density-field information quantified by the clustering multipoles.  Analyses including the redshift-space distortion signal contain more information (since they are additionally sensitive to the normalised growth of structure parameter $f \sigma_8$), but bring significant additional complexity in modelling the redshift-space clustering.

We performed the shear correlation function measurements using {\tt treecorr}{\footnote{\url{https://rmjarvis.github.io/TreeCorr}}} \citep{2004MNRAS.352..338J}, adopting the same angular separation bins as utilised by the different weak lensing collaborations in their main cosmological analyses.  Hence, for KiDS-1000 we use 9 logarithmic bins in the angular range $0.5 < \theta < 300$ arcmin \citep{2021A&A...645A.104A}, for DES-Y3 we use 20 bins in the range $2.5 < \theta < 250$ arcmin \citep{2022PhRvD.105b3514A, 2022PhRvD.105b3515S}, and for HSC-Y1 we use 22 bins in the range $10^{0.05} < \theta < 10^{2.25}$ arcmin \citep{2020PASJ...72...16H} (where for cosmological analysis we apply the same angular scale cuts as adopted by each collaboration).  For the average tangential shear measurements we also used {\tt treecorr}, employing 15 logarithmic angular separation bins in the range $0.3 < \theta < 300$ arcmin.  We measured the projected correlation functions using {\tt corrfunc}{\footnote{\url{https://corrfunc.readthedocs.io}}} \citep{2020MNRAS.491.3022S} with 15 logarithmic projected separation bins in the range $0.08 < R < 80 \, h^{-1}$ Mpc (noting that for cosmological analysis we impose a minimum projected separation scale when fitting models to these statistics).  We set the maximum radial separation in the projected correlation function measurement as $\Pi_\textrm{max} = 100 \, h^{-1}$ Mpc.

The galaxy-galaxy lensing and clustering correlation measurements also utilised random lens catalogues, matching the angular and redshift distributions of the data lenses.  We used the weights assigned to the weak lensing sources (which resemble the weights of the real weak lensing datasets as discussed in Sec.~\ref{sec:buzzard}), and do not weight the DESI galaxies (since the simulations do not include incompleteness effects).  We corrected the shear and galaxy-galaxy lensing correlations for the multiplicative shear bias introduced in the simulations.

\begin{figure*}
\centering
\includegraphics[width=1.8\columnwidth]{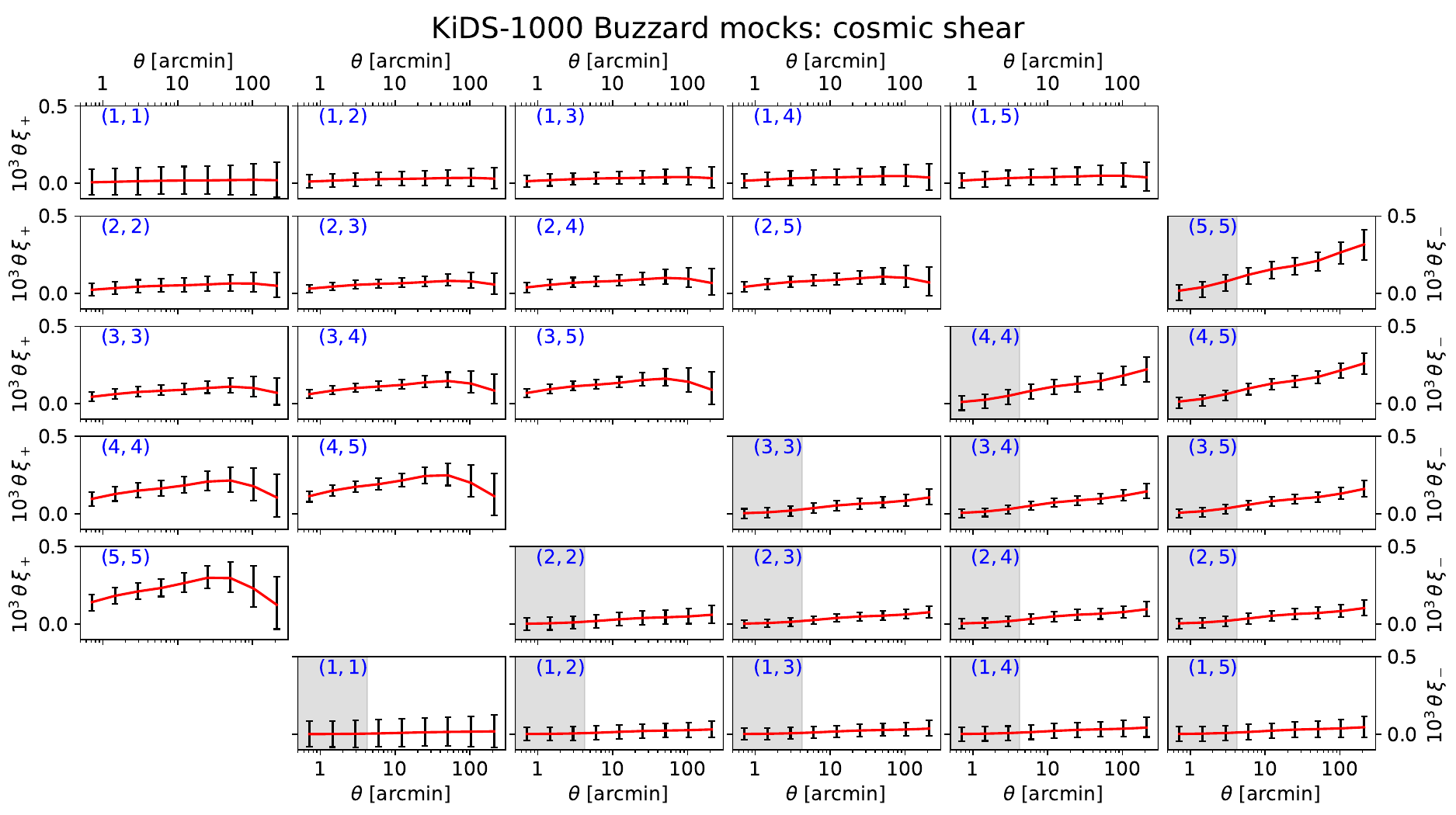}
\caption{The cosmic shear correlation functions $(\xi_+, \xi_-)$ between the five tomographic source samples of the KiDS-1000 Buzzard mocks, as a function of angular separation $\theta$ in arcminutes.  The different panels display measurements for different combinations of tomographic samples, as indicated by the identifiers in the top-left corner of each panel.  We plot the mean and standard deviation of the measurements across 160 Buzzard regions.  Shaded sections of panels indicate scale ranges excluded from cosmological analysis by the KiDS collaboration.  The solid lines show the fiducial cosmological models for $(\xi_+, \xi_-)$ and the $y$-axes are scaled by a factor $10^3 \times \theta$ (with $\theta$ in arcminutes) for clarity of presentation.}
\label{fig:xipm_buzzard_kids}
\end{figure*}

We illustrate our results using the example of the KiDS-1000-DESI simulations, noting that the analyses of the DES-Y3 and HSC-Y1 configurations reached similar conclusions.  Fig.~\ref{fig:xipm_buzzard_kids} displays shear correlation function measurements, $\xi_\pm(\theta)$, between the five tomographic source samples of the KiDS-1000 Buzzard mocks.  We plot the mean and standard deviation of the measurements for the different regions, compared to the fiducial cosmological models in each case (introduced in Sec.~\ref{sec:model}).  The average difference between the mock mean and the model, relative to the forecast error, is less than $0.05\sigma$ for both $\xi_+$ and $\xi_-$.

\begin{figure*}
\centering
\includegraphics[width=1.8\columnwidth]{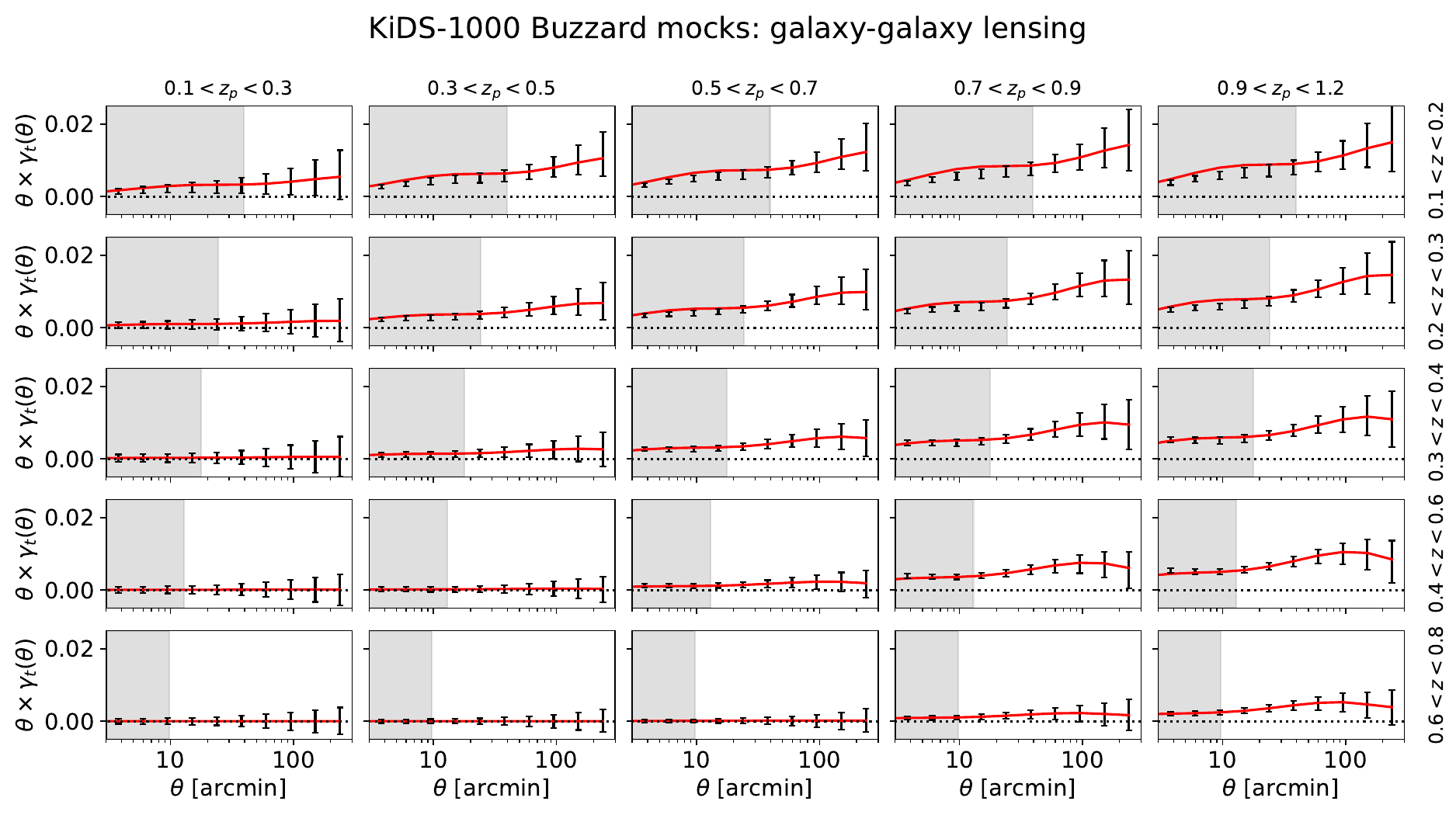}
\caption{The average tangential shear $\gamma_t$ of the five tomographic source samples of the KiDS-1000 Buzzard mocks, around the five lens samples of the DESI Buzzard mocks, as a function of angular separation $\theta$ in arcminutes.  The different panels display measurements for different combinations of source and lens samples, as indicated by the captions above and to the right of each panel.  We plot the mean and standard deviation of the measurements across 160 Buzzard regions.  Shaded sections of panels indicate scale ranges excluded from cosmological analysis by an example cut $R < 5 \, h^{-1}$ Mpc, where $R$ is the projected co-moving scale corresponding to angular separation $\theta$ at the mid-point of each lens redshift bin.  The solid lines show the fiducial cosmological models for $\gamma_t$, including a linear bias factor for each lens sample, and the $y$-axes are scaled by a factor $\theta$ (in arcminutes) for clarity of presentation.}
\label{fig:gt_buzzard_kids}
\end{figure*}

Fig.~\ref{fig:gt_buzzard_kids} displays the $\gamma_t(\theta)$ measurements between the five tomographic source samples of the KiDS-1000 Buzzard mocks, around the five lens samples of the DESI Buzzard mocks.  We plot the mean and standard deviation of the measurements for the different regions, compared to the fiducial cosmological models in each case, scaled by a linear galaxy bias factor at each lens redshift.  The average difference between the mock mean and the model, relative to the forecast error, is less than $0.1\sigma$ in the range of projected separations $R > 5 \, h^{-1}$ Mpc where a linear galaxy bias model might be potentially applicable (we will test the fitting ranges for cosmological analysis below).

\begin{figure*}
\centering
\includegraphics[width=1.8\columnwidth]{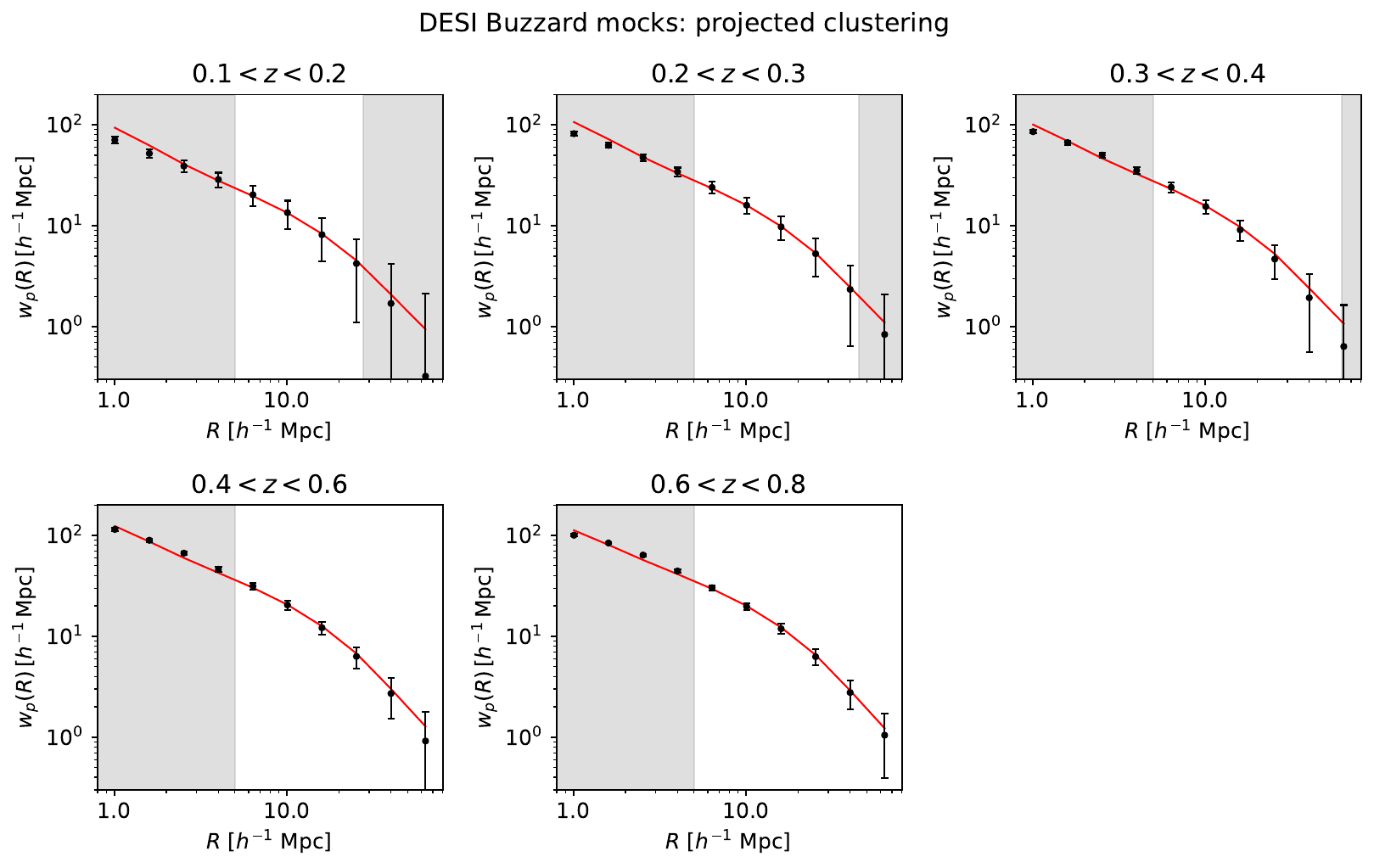}
\caption{The projected correlation function $w_p$ of the five lens samples of the DESI Buzzard mocks, as a function of projected separation $R$ in units of $h^{-1}$ Mpc.  We plot the mean and standard deviation of the measurements across the Buzzard KiDS regions.  Shaded sections of panels indicate scale ranges excluded from cosmological analysis by an example cut $R < 5 \, h^{-1}$ Mpc (on small scales), and a conservative large-scale cut we include to remove any effect of the domain decomposition adopted when constructing the mock galaxy catalogues.  The solid lines show the fiducial cosmological models.}
\label{fig:wp_buzzard_kids}
\end{figure*}

Fig.~\ref{fig:wp_buzzard_kids} displays projected correlation function measurements of the five lens samples of the DESI Buzzard mocks.  We plot the mean and standard deviation of the measurements for the different regions, compared to the fiducial cosmological models in each case.  The principal deviations between the measurements and model are on small scales, where the linear bias model is not applicable and Buzzard mocks are not fully resolved, and to a significantly lesser degree on large scales, where the clustering is impacted by the domain decomposition used when constructing the mocks as discussed by \citealt{2019arXiv190102401D} and \citealt{2022ApJ...931..145W}.  When populating galaxies into lightcones, {\tt ADDGALS} splits volumes into $n_\text{side}=4$ domains. The number of galaxies populated in each of these domains is modulated by the dark matter over density averaged over the domain, and thus the galaxy density fluctuations are artificially impacted on scales larger than the size of the domain making the galaxy bias relation unreliable on these scales.  We also note in Fig.~\ref{fig:wp_buzzard_kids} that the poorer agreement between the mock mean and models in the $0.3 < z < 0.4$ lens bin might be caused by the transition between the two light cones used to construct the Buzzard simulations, as noted in Sec.~\ref{sec:simulations}.  In Sec.~\ref{sec:modelfits} we will consider full cosmological fits to these measurements.

\section{Combined-probe covariance}
\label{sec:covariance}

\subsection{Analytical covariance}
\label{sec:anacov}

We used analytical methods to determine the covariances between different statistics, redshifts and scales for the DESI samples and the DES, KiDS and HSC weak lensing datasets.  We focus on the covariances amongst the $3 \times 2$-pt correlation functions comprised of the cosmic shear, $\xi_\pm(\theta)$, the average tangential shear, $\gamma_t(\theta)$, and the projected correlation function, $w_p(R)$.  Future analyses will extend the joint covariance to the multipoles of the clustering correlation function, including redshift-space distortions.

Our covariance computation follows the methods of \cite{2017MNRAS.470.2100K} and \cite{2021A&A...646A.129J}, with some modifications to treat the case of $w_p(R)$, and is fully described in Appendices \ref{sec:gaussian}, \ref{sec:supersample}, \ref{sec:noisecorr} and \ref{sec:covwp}.  Appendix \ref{sec:gaussian} details the computation of a Gaussian analytical covariance including the sample variance, noise and mixed contributions.  This Gaussian part of the covariance assumes an effective survey area for each component, rather than considering the full survey mask.  Appendix \ref{sec:supersample} describes the calculation of the super-sample covariance we include in our computation, which does use the survey mask.  Appendix \ref{sec:noisecorr} specifies the correction of the noise covariance by the measured number of survey pairs, and Appendix \ref{sec:covwp} details how the covariances are modified for the projected correlation function.  We do not include the non-Gaussian contribution to the covariance, which is currently negligible for the cases of interest \citep{2021A&A...646A.129J}.  Appendix \ref{sec:codecomparison} displays comparisons of our analytical covariance evaluations to those of two existing codes: {\tt CosmoCov} \citep{2017MNRAS.470.2100K,2020MNRAS.497.2699F}, and the code used in the KiDS-1000 cosmology analysis \citep{2021A&A...646A.129J}\footnote{We thank Benjamin Joachimi for sharing the analytical covariance code for the purpose of conducting this test.  We note that since we completed our analysis, the KiDS covariance code has been publicly released and is presented by \cite{2024arXiv241006962R}.}.  The evaluations agree to $1\%$ accuracy for equivalent configurations.

\subsection{Covariance tests}

In this section we test the accuracy of the analytical covariance determinations for the different survey configurations, correlation functions and scales, through comparison with the ensemble of Buzzard mock measurements.  We again illustrate our results using the example of the KiDS-1000-DESI configuration, noting that the performance is comparable for the DES-Y3 and HSC-Y1 configurations.

\begin{figure*}
\centering
\includegraphics[width=1.8\columnwidth]{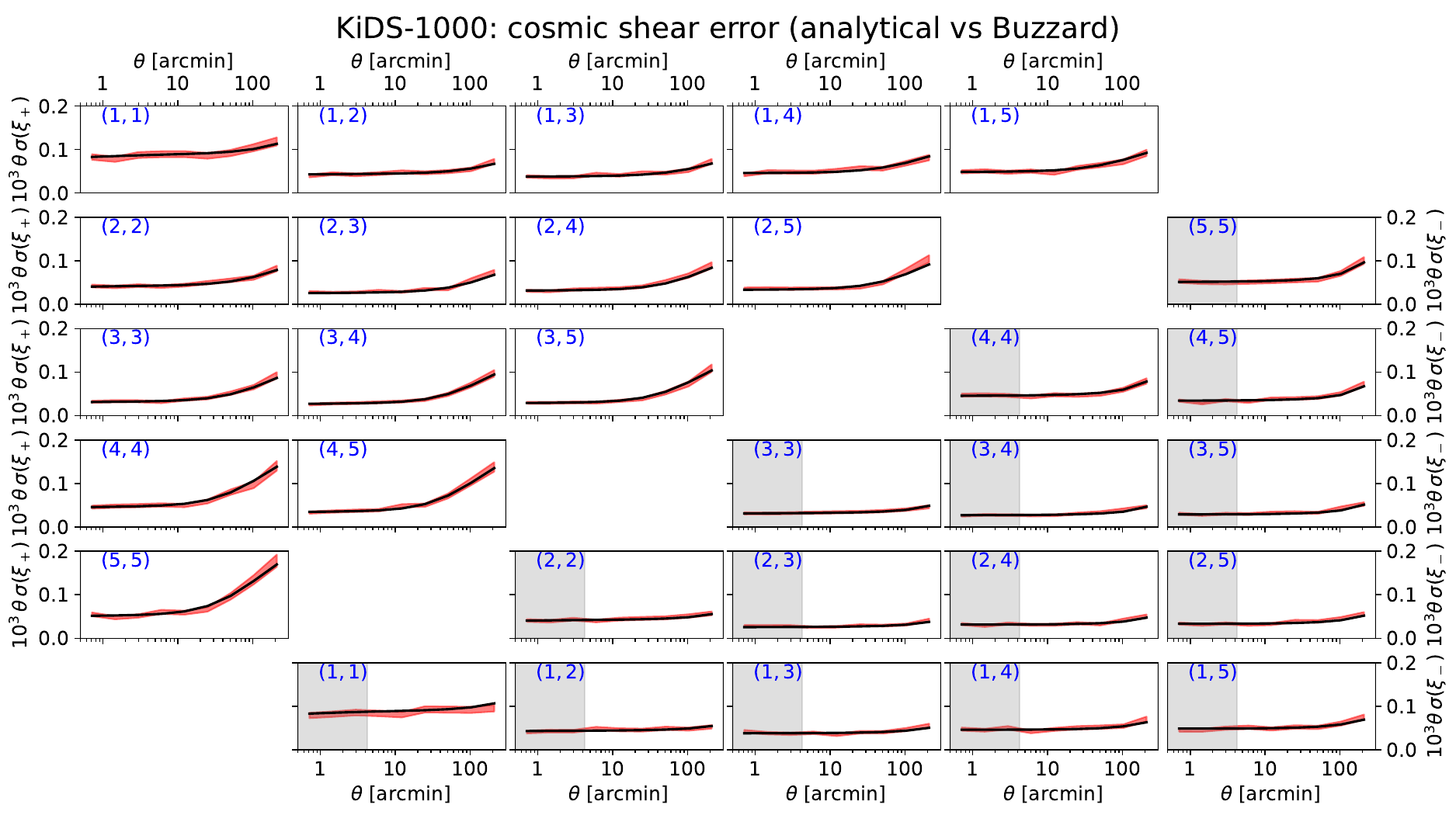}
\caption{The error in the shear correlation function measurements $(\xi_+, \xi_-)$, denoted $\sigma(\xi_+)$ and $\sigma(\xi_-)$, between the five tomographic source samples of the KiDS-1000 Buzzard mocks, as a function of angular separation $\theta$ in arcminutes.  The different panels display measurements for different combinations of tomographic samples, as indicated by the identifiers in the top-left corner of each panel.  We compare the measurement error in each separation bin predicted by the analytical covariance (black line), with the standard deviation of the measurements across 160 Buzzard regions (the red band, which indicates the error in the inferred standard deviation arising from the limited number of regions).  The $y$-axes are scaled by a factor $10^3 \times \theta$ in arcminutes, for clarity of presentation.  Shaded sections of panels indicate scale ranges excluded from cosmological analysis by the KiDS collaboration.}
\label{fig:xipmerr_buzzard_kids}
\end{figure*}

Fig.~\ref{fig:xipmerr_buzzard_kids} shows the comparison of the error in the $\xi_\pm(\theta)$ measurements for all combinations of the five KiDS-1000 tomographic source samples, between the analytical covariance and the standard deviation of measurements across the Buzzard regions.  As described in Sec.~\ref{sec:anacov}, the analytical covariance includes the Gaussian and super-sample contributions, together with the noise correction using the average number of observed pairs.  The mock measurements are displayed as a band, where the width of the band indicates the ``error in the error'' appropriate for Gaussian statistics, $\Delta \sigma \approx \sigma/\sqrt{2 N_{\rm mock}}$, where $N_{\rm mock} = 160$ in the case of the KiDS-DESI simulations.  We find that the analytical covariance provides an accurate model for the dispersion between mocks.  The average difference between the analytical and mock error in each separation bin, across the different measurements, is less than $1\%$ for both $\xi_+$ and $\xi_-$.  The standard deviation of the differences is less than $5\%$ (using all separations), which is consistent with the error in the numerical estimate of the covariance.

\begin{figure*}
\centering
\includegraphics[width=1.8\columnwidth]{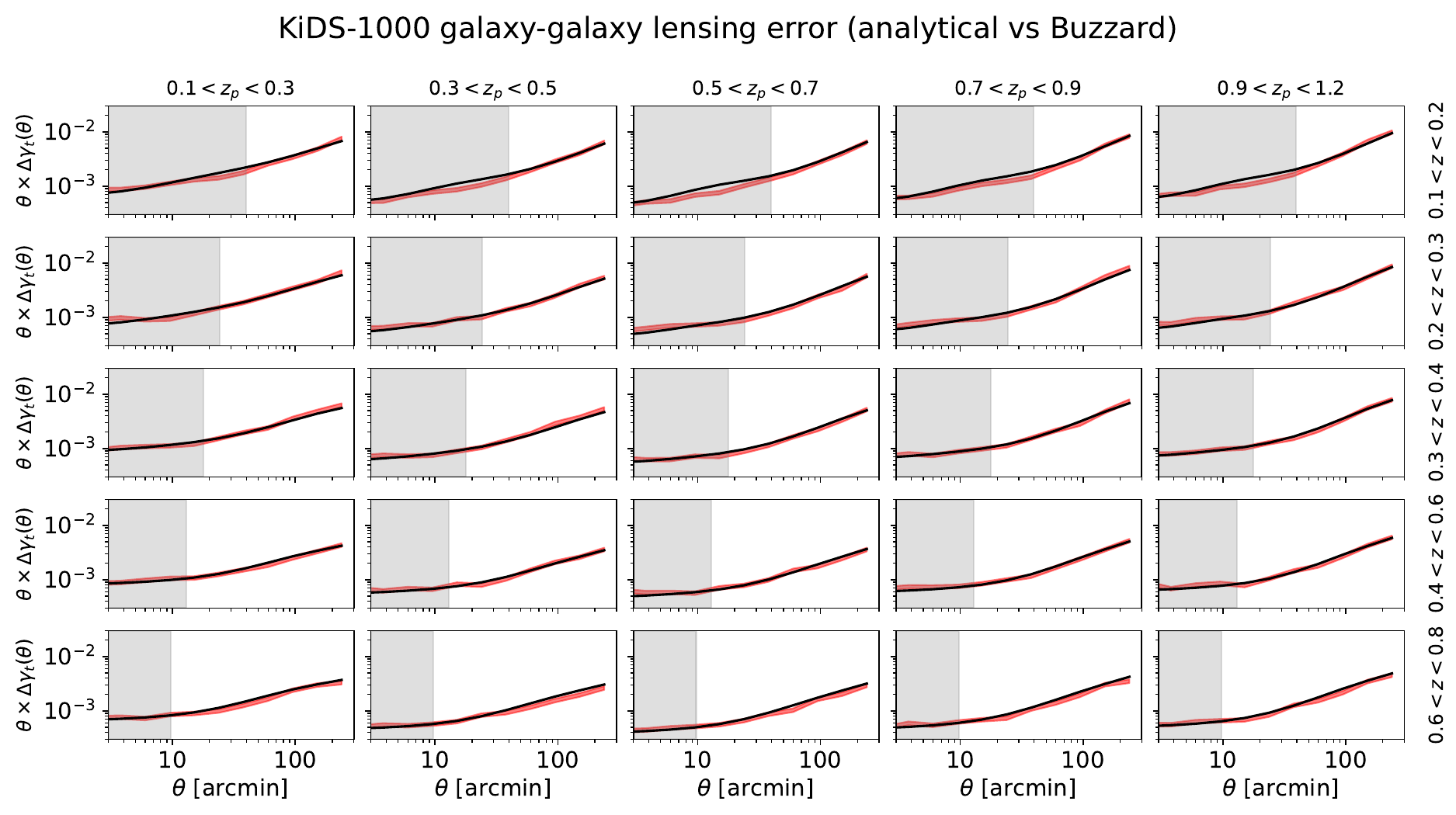}
\caption{The error in the average tangential shear $\gamma_t$ of the five tomographic source samples of the KiDS-1000 Buzzard mocks, around the five lens samples of the DESI Buzzard mocks, as a function of angular separation $\theta$ in arcminutes.  The different panels display measurements for different combinations of source and lens samples, as indicated by the captions above and to the right of each panel.  We compare the measurement error in each separation bin predicted by the analytical covariance (black line), with the standard deviation of the measurements across 160 Buzzard regions (the red band, which indicates the error in the inferred standard deviation arising from the limited number of regions).  Shaded sections of panels indicate ranges excluded from cosmological analysis by a cut $R < 5 \, h^{-1}$ Mpc.}
\label{fig:gterr_buzzard_kids}
\end{figure*}

Next, we analyse the performance of the analytical covariance of the average tangential shear measurements, $\gamma_t(\theta)$.  Fig.~\ref{fig:gterr_buzzard_kids} shows a comparison of the error in the $\gamma_t(\theta)$ measurements, between the analytical covariance and the standard deviation of measurements across the Buzzard regions.  The results are displayed in a similar format to Fig.~\ref{fig:xipmerr_buzzard_kids}.  The average difference between the analytical and mock error in each separation bin, across the different measurements, is less than $5\%$ (using all separations), and the standard deviation of the differences is around $10\%$.  The principal deviations are close to the 1-halo transition scale where, as evidenced in Fig.~\ref{fig:gt_buzzard_kids}, the linear bias model does not accurately represent the average tangential shear signal.  These scales are excluded from our analysis.  For the projected correlation function $w_p(R)$, the domain decomposition used in the construction of the Buzzard galaxy catalogues implies that our estimate of the numerical covariance is less reliable and we could not perform an analogous test, but we will assess the goodness-of-fit of analyses using these statistics in Sec.~\ref{sec:modelfits} below.

\begin{figure*}
\centering
\includegraphics[width=1.8\columnwidth]{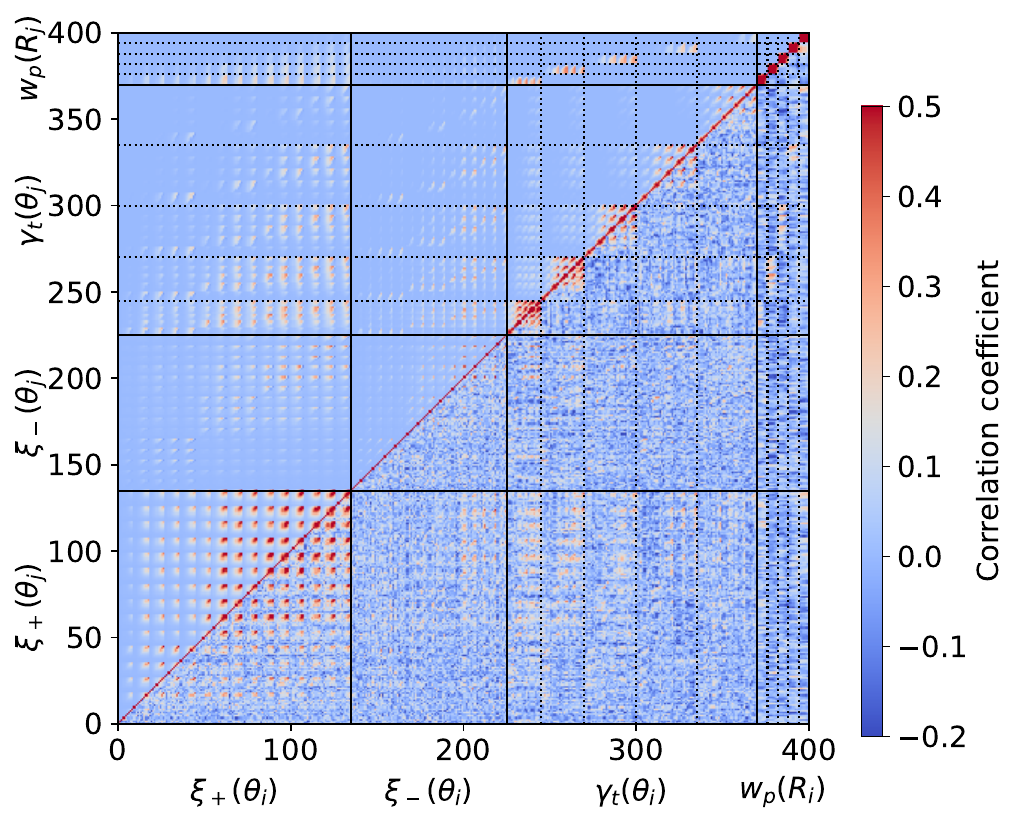}
\caption{The correlation matrix corresponding to the combined-probe covariance of the 3$\times$2-pt correlation function measurement of the KiDS-1000 and DESI Buzzard mocks.  The measurements use $N_{\rm tom} = 5$ source tomographic samples and $N_{\rm lens} = 5$ lens samples.  The analytical covariance is depicted in the upper-left triangle, and the numerical covariance derived from 160 realisations in the lower-right triangle.  The data vector is ordered such that the different elements of the $\xi_+$, $\xi_-$, $\gamma_t$ and $w_p$ correlation functions are grouped together.  This allows the correlation matrix to be divided into sections corresponding to different statistics, indicated by the solid horizontal and vertical lines.  The $\gamma_t$ and $w_p$ statistics are also divided into $N_{\rm lens} = 5$ lens samples, indicated by the dotted lines.  We have applied separation cuts as explained in the text, reducing the size of the original covariance matrix.}
\label{fig:covcut_xigtwp_kids}
\end{figure*}

\begin{figure}
\centering
\includegraphics[width=\columnwidth]{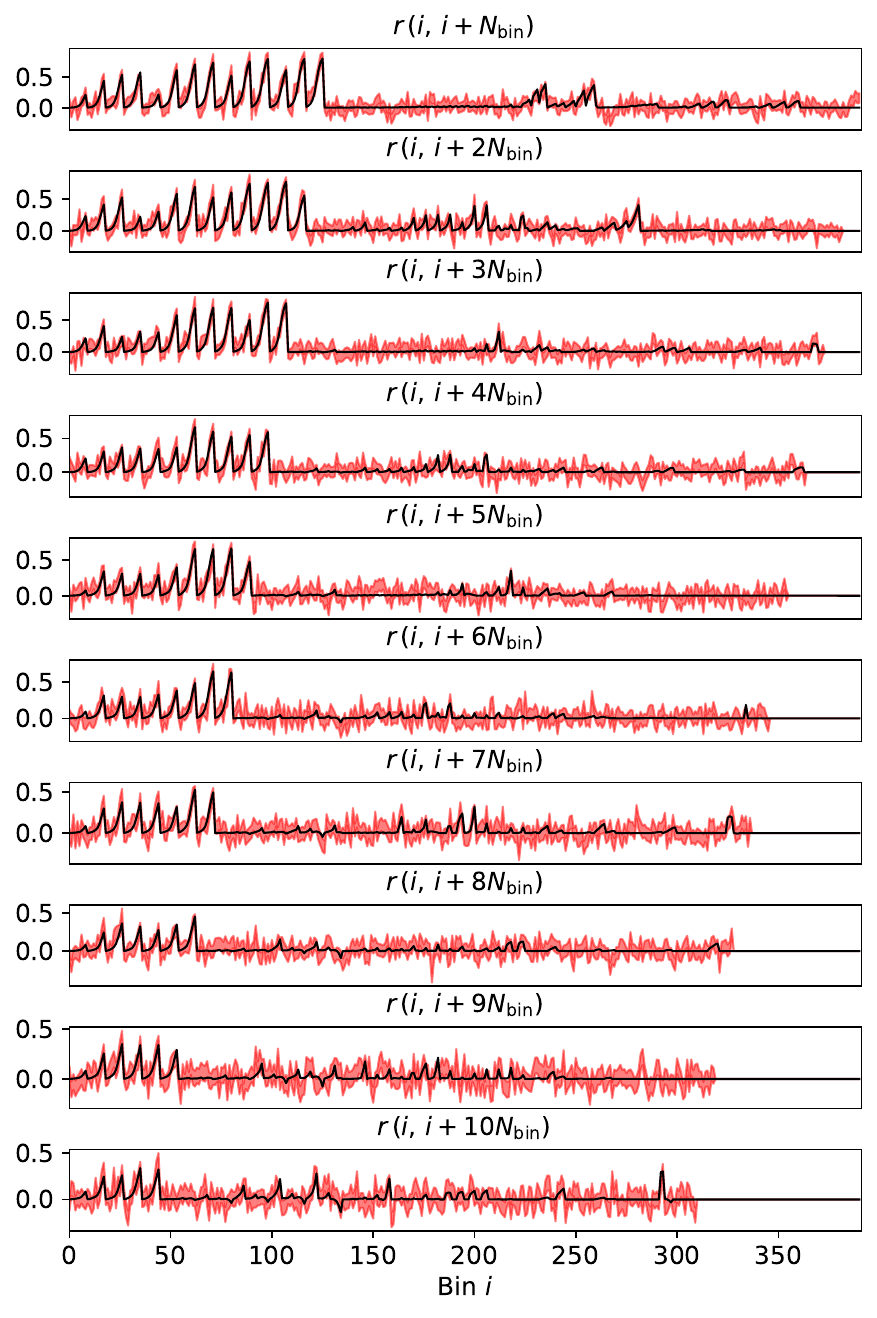}
\caption{The correlation coefficient of the off-diagonal combined-probe covariance matrix corresponding to KiDS-1000 and the DESI Buzzard mocks, $r(i, j) = C_{ij}/\sqrt{C_{ii} \, C_{jj}}$. The different panels display the full set of correlation coefficients for matrix elements sharing the same scale bin, for different lens and source bins offset from the central diagonal, where $N_{\rm bin} = 9$ for the KiDS cosmic shear measurements.  We compare the correlation coefficient predicted by the analytical covariance (solid black line) and the Buzzard realisations (red band).}
\label{fig:offdiagcorr}
\end{figure}

Fig.~\ref{fig:covcut_xigtwp_kids} depicts the full structure of the analytical covariance matrix of the 3$\times$2-pt correlation function measurement of the KiDS-1000 and DESI Buzzard mocks as a correlation matrix, where the analytical covariance is displayed in the upper-left triangle, and the numerical covariance in the lower-right triangle.  Different subsets of the data vector, corresponding to the $\xi_+$, $\xi_-$, $\gamma_t$ and $w_p$ correlation functions, are delineated by the solid horizontal and vertical lines.  For this representation, we have applied separation cuts to the data vector, reducing the size of the covariance matrix.  For the shear correlation functions we use the same separation cuts as the weak lensing collaborations, and for the galaxy-galaxy lensing and clustering correlation functions we apply a cut in projected separation $R > 5 \, h^{-1}$ Mpc in each lens redshift bin (for visualisation purposes, where we perform a quantitative assessment of the scale cuts in Sec.~\ref{sec:modelfits}).  In the case of the KiDS-DESI configuration, there are $(135, 90, 145, 30)$ elements remaining for $(\xi_+, \xi_-, \gamma_t, w_p)$ after these cuts, hence the size of the reduced covariance matrix is $400 \times 400$ elements.  Fig.~\ref{fig:offdiagcorr} displays the amplitude of the cross-correlation coefficient in several off-diagonal cuts through the covariance matrix.  Whilst the analytical and numerical estimates of the covariance matrix are consistent, we acknowledge that the number of mock realisations is not sufficient to accurately validate a small off-diagonal signal.

\section{Model fitting}
\label{sec:modelfits}

In this section we use the simulated measurements and corresponding analytical covariance to create combined-probe cosmological parameter fits resembling those which will be applied using the DESI-Y1 dataset.

\subsection{Data vectors}
\label{sec:datavectors}

The combined-probe analysis we wish to simulate consists of weak lensing and clustering datasets with different sky coverage, which partially overlap in a region in which galaxy-galaxy lensing may be measured.  In order to better represent this configuration in our tests, we extended the $\xi_\pm(\theta)$ and $w_p(R)$ measurements presented in Sec.~\ref{sec:simulations} beyond the overlap region traced by $\gamma_t(\theta)$ measurements.  Allowing for the extended area required, we are only able to create a single simulated data vector from each of the Buzzard quadrants for these tests.

We constructed the clustering portion of the data vector from a simulation area of $4833$ deg$^2$ (that is, 90 pixels corresponding to $n_{\rm side} = 8$).  This is representative of the areal coverage of DESI-Y1 noting that, although the Y1 dataset has partial coverage over a larger area, we do not consider this varying completeness in our analysis.  We created the weak lensing survey datasets from simulation areas of $(967, 4028, 161)$ deg$^2$ for (KiDS-1000, DES-Y3, HSC-Y1), which are comprised of $(18, 75, 3)$ Buzzard pixels (which are convenient choices because they correspond to constant multiples of the regions we defined in Sec.~\ref{sec:simulations})\footnote{The effective areas of the actual weak lensing datasets can be estimated as $773$ deg$^2$ for KiDS-1000 \citep{2021A&A...646A.129J}, $4143$ deg$^2$ for DES-Y3 \citep{2021MNRAS.504.4312G} and $137$ deg$^2$ for HSC-Y1 \citep{2018PASJ...70S..25M}.  Our simulations are representative of the configurations of these surveys, but do not match them exactly.}.  The overlap areas of these simulated datasets match those chosen in Sec.~\ref{sec:simulations}.

We adapted the covariance determinations described in Sec.~\ref{sec:covariance} to allow for the different areal coverage of the components of the $3 \times 2$-pt data vector.  Suppose the areal coverage of the datasets used in two measurements is $\Omega_1$ and $\Omega_2$, and $\Omega_O$ is the overlap area common to both datasets.  We can split correlation function measurements for both datasets into two pieces corresponding to the non-overlapping (``$NO$'') and overlapping (``$O$'') components, weighted by area as:
\begin{equation}
\begin{split}
  \xi_1(\theta) &= \frac{(\Omega_1 - \Omega_O) \, \xi_{1,NO}(\theta) + \Omega_O \, \xi_{1,O}(\theta)}{\Omega_1} , \\
  \xi_2(\theta) &= \frac{(\Omega_2 - \Omega_O) \, \xi_{2,NO}(\theta) + \Omega_O \, \xi_{2,O}(\theta)}{\Omega_2} ,
\end{split}
\end{equation}
where $\xi$ can represent any cosmic shear, galaxy-galaxy lensing or angular clustering correlation function.  Combining these equations and assuming no correlation between the disjoint regions \citep[which was found to be a safe assumption by][]{2023OJAp....6E..36D} we find,
\begin{equation}
  {\rm Cov} \left[ \xi_1(\theta) , \xi_2(\theta') \right] = \frac{\Omega_O^2}{\Omega_1 \, \Omega_2} \, {\rm Cov} \left[ \xi_{1,O}(\theta) , \xi_{2,O}(\theta') \right] .
\end{equation}
Hence, partial sky overlap can be corrected by scaling the covariance determined for the common area by the pre-factor $\Omega_O^2/\Omega_1 \Omega_2$.

\subsection{Model description and parameterisation}
\label{sec:model}

In this section we describe the models we adopt for the likelihood analysis of these data vectors.  The full set of parameters we use in our analysis is listed in Table \ref{tab:priors}.  We sample over five base cosmological parameters: the matter density $\Omega_m$, baryon density $\Omega_b$, Hubble constant $H_0$, spectral index $n_s$, and the amplitude of the primordial power spectrum, $A_s$.  We will express the amplitude fits in terms of two alternative late-time parameters: the normalisation of the $z=0$ matter power spectrum, $\sigma_8$, and the standard parameter combination $S_8 = \sigma_8 \sqrt{\Omega_m / 0.3}$ that is best constrained by cosmic shear datasets.

\subsubsection{Cosmic shear}

We calculate the cosmic shear model for the $i$-th and $j$-th tomographic bins using the flat-sky approximation as,
\begin{equation}
    \xi^{ij}_{+/-}(\theta) = \int \frac{d\ell \, \ell}{2\pi} \frac{J_{0/4}(\ell\theta)}{1+\ell/\ell_\text{max}} \, C_{\kappa\kappa}^{ij}(\ell) ,
\label{eq:ximod}
\end{equation}
where $J_0$ and $J_4$ are the $0$th and $4$th order Bessel functions of the first kind, respectively. The $\ell$-dependent factor in the denominator of Eq.~\ref{eq:ximod} ensures that the modelling of the cosmic shear correlation functions does not extend to scales smaller than those valid for the ray-tracing algorithm given the finite pixel size, following \cite{2019arXiv190102401D}, although we note that the exclusion of this factor would not significantly affect our results.  We use $\ell_\text{max} = \pi/\theta_\text{pix}$, where $\theta_\text{pix} = 0.46^\prime$ is the effective pixel size, and therefore we only consider scales where $\ell < \ell_\text{max}$.

The angular convergence power spectrum, $C_{\kappa\kappa}^{ij}(\ell)$ in Eq.~\ref{eq:ximod}, is given in the Limber approximation by \citep[e.g.,][]{2001MNRAS.321..439G, 2004PhRvD..70d3009H, 2010A&A...523A...1J},
\begin{equation}
    C_{\kappa\kappa}^{ij} = \int \frac{d\chi}{\chi^2} \, P_{mm}\left(\frac{\ell+\frac{1}{2}}{\chi},z(\chi)\right) W^i_{\kappa,s}(\chi) W^j_{\kappa,s}(\chi) ,
\end{equation}
where the lensing efficiency kernel is,
\begin{equation}
    W^i_{\kappa,s}(\chi) = \frac{3\Omega_m H_0^2}{2c^2} \frac{\chi}{a(\chi)} \int_\chi^{\chi_\text{h}} d\chi^\prime n_s^i(\chi^\prime) \frac{(\chi^\prime - \chi)}{\chi^\prime} ,
    \label{eq:lensing_kernel}
\end{equation}
and $P_{mm}(k,z)$ is the matter power spectrum as a function of wavenumber $k$ and redshift $z$.  In Eq.~\ref{eq:lensing_kernel}, the label $s$ represents the sources, $n_s^i$ corresponds to the redshift distribution of the source galaxies in tomographic bin $i$, $H_0$ represents the Hubble constant, $c$ is the speed of light in vacuum and $a$ is the scale factor.  We calculate the matter power spectrum $P_{mm}(k,z)$ for a set of cosmological parameters using {\tt camb}~\citep{2000ApJ...538..473L}, adopting the {\tt halofit} model~\citep{2012ApJ...761..152T} that was shown to successfully reproduce the cosmic shear measurements from the Buzzard simulation suite~\citep{2019arXiv190102401D}. We note that {\tt halofit} may not be sufficiently accurate for future analyses \citep[as discussed by, e.g.,][]{2021MNRAS.502.1401M, 2023JCAP...07..054D}, and we discuss this issue further in \cite{NimasDESI} and \cite{AnnaDESI}, which present the modelling choices for the real data analysis.

\subsubsection{Galaxy-galaxy lensing}

Under the flat-sky approximation we can compute the average tangential shear correlation function as,
\begin{equation}
    \gamma_t^{ij}(\theta) = \int \frac{d\ell \, \ell}{2\pi} \, J_2(\ell\theta) \, C_{g\kappa}^{ij}(\ell) ,
\end{equation}
where $J_2$ is the $2$nd order Bessel function of the first kind.  For the galaxy-convergence cross-power spectrum, $C_{g\kappa}^{ij}(\ell)$, we consider contributions from galaxy density and magnification transfer functions such that in the Limber approximation,
\begin{equation}
\begin{split}
    C_{g\kappa}^{ij} = &\int \frac{d\chi}{\chi^2} \, P_{mm}\left(\frac{\ell+\frac{1}{2}}{\chi},z(\chi)\right) \\ & \times \left[ W^i_{\delta,l}(\chi)+W^i_{\mu,l}(\chi) \right] \, W^j_{\kappa,s}(\chi) .
\end{split}
\end{equation}
The galaxy density kernel in the lens redshift bin $i$ is,
\begin{equation}
    W^i_{\delta,l}(\chi) = b_1^i \, n_l^i(z) \, \frac{dz}{d\chi},
\end{equation}
where we have assumed a linear galaxy bias parameter $b_1^i$ per redshift bin, and $n_l^i$ is the lens redshift distribution for the $i$-th bin. We include the contributions from lensing magnification following \cite{2023MNRAS.523.3649E} and calculate the lens magnification efficiency kernel as,
\begin{equation}
    W^i_{\mu,l}(\chi) = 2(\alpha_l^i-1) W^i_{\kappa,l}(\chi) .
\end{equation}
Thus, the contribution from lens magnification to $\langle \delta_g^\text{obs}\gamma \rangle$ is simply $2(\alpha_l^i-1)\langle \kappa_l\gamma \rangle$, in agreement with \cite{2022PhRvD.105h3528P}.

\subsubsection{Projected correlation function}

We analyse the projected clustering correlation function $w_p(R)$ of the lenses as a function of projected separation $R$.  First, we calculate the galaxy power spectrum in redshift space assuming a linear Kaiser model~\citep{1987MNRAS.227....1K},
\begin{equation}
    P_{gg}^s(k, \mu, z_l) = P_{mm}(k,z_l) \, \left[ b(z_l,k) + f(z_l)\mu^2 \right]^2, 
\label{eq:pggmod}
\end{equation}
where $b(z_l,k)$ is the galaxy bias, $f(z_l)$ is the growth rate, and $\mu = \cos(\theta)$ in terms of the angle $\theta$ of the clustering wavevector with respect to the line-of-sight.  We assume a linear galaxy bias parameter for each lens galaxy sample.  We can adopt the simple galaxy clustering model shown in Eq.\ref{eq:pggmod} because the projected correlation function is not sensitive to redshift-space distortions, and we're restricting our analysis to large scales.

We then expand the galaxy power spectrum into Legendre multipoles as,
\begin{equation}
    P_\ell^s(k) = \frac{2\ell+1}{2}\int^{1}_{-1} P^s_{gg}(k, \mu) \, L_\ell(\mu) \, d\mu ,
\end{equation}
where $L_\ell(\mu)$ are the Legendre polynomials. The correlation function multipoles in redshift space can be calculated from the power spectrum multipoles by a Hankel transform as,
\begin{equation}
    \xi_\ell^s(s) = i^\ell \int_0^\infty \frac{dk \, k^2}{2\pi^2} \, P_\ell^s(k) \, j_\ell(ks) ,
\end{equation}
where $j_\ell$ are the $\ell$-th order spherical Bessel functions. We calculate the Hankel transform using the {\tt FFTlog-and-beyond}\footnote{\url{https://github.com/xfangcosmo/FFTLog-and-beyond}} package from \cite{2020JCAP...05..010F} which solves transformations of the form $\int_0^\infty \frac{dx}{x}f(x)j_\ell^{(n)}(xy)$.

We can compute the projected correlation function from the galaxy correlation function multipoles as \citep[see, e.g.,][]{2010PhRvD..81f3531B, 2021MNRAS.501.4167S},
\begin{equation}
\begin{split}
    w_p(R) = 2\sum_{\ell=\{0,2,4\}} & \int_0^{\Pi_\text{max}} d\Pi \, \xi_{\ell}^s\left(\sqrt{R^2+\Pi^2}\right) \\ & \times L_{\ell}\left( \frac{\Pi}{\sqrt{R^2+\Pi^2}}\right) ,
\end{split}
\end{equation}
where $\Pi$ is the line-of-sight separation and $\Pi_\text{max}$ is set to be $100 \, h^{-1}$ Mpc.  We only use the monopole, quadrupole and hexadecapole terms of the multipole expansion in our calculation of $w_p(R)$, although we note that the contribution of the hexadecapole is negligible by comparison with the monopole and quadrupole.

When evaluating $w_p(R)$ we rescale $R$ and $\Pi$ to take into account the Alcock-Paczynski effect (AP)~\citep{1979Natur.281..358A}, that introduces distortions when converting redshifts to distances using a trial cosmological model which differs from the fiducial model in which the measurements have been performed. We include the AP effect following \cite{2017A&A...608A..44D}, and re-scale our distances as
\begin{equation}
    R' = R \left[ \frac{D_A^{\text{trial}}(z_l)}{D_A^{\text{fid}}(z_l)} \right]
\end{equation}
and
\begin{equation}
    \Pi' = \Pi \left[ \frac{H^{\text{fid}}(z_l)}{H^{\text{trial}}(z_l)} \right].
\end{equation}
Here, the label ``trial'' corresponds to a trial cosmology which is being assumed at each step of the chain and ``fid'' stands for the fiducial cosmology in which the measurements have been originally performed, which in our case corresponds to that of the Buzzard simulation suite.

\subsubsection{Systematics and priors}
\label{subsubsection:nuisance_and_priors}

In this section we describe the nuisance parameters that we include in our parameter fit to account for astrophysical and measurement systematics in the cosmological fits.

\textit{Galaxy bias.} As the lens galaxies are biased tracers of the underlying mass density field, we include galaxy bias parameters. As mentioned above, we adopt a linear galaxy bias model in each redshift bin as,
\begin{equation}
    b^i(z,k) = b_1^i ,
\end{equation}
where $b(z,k)$ is the general bias form included in Eq.~\ref{eq:pggmod}. We hence include a total of five galaxy bias parameters in our fits, one for each lens redshift bin.  We note that our model excludes non-linear galaxy formation effects such as assembly bias and segregation of satellite galaxies relative to dark matter \citep{2023MNRAS.521..937C, 2023MNRAS.525.3149C}.

\textit{Photometric redshift uncertainties.}  We account for uncertainties in the redshift distributions of the source galaxies by introducing shift parameters according to,
\begin{equation}
    n_s^i(z) = n_s^i(z-\Delta z_s^i) ,
\end{equation}
replicating how these uncertainties are treated in standard lensing cosmological analyses.  We hence marginalize over a total of four shift parameters for DES-Y3 and HSC-Y1, given the division of those samples into four tomographic source bins, whilst we have five parameters to marginalize for KiDS-1000. Since the redshifts of our lens galaxy sample are obtained using spectroscopic techniques, their uncertainties are negligible.

\textit{Multiplicative shear bias.}  Following the analysis of real lensing datasets, we include multiplicative shear bias parameters $m_i$ in our cosmic shear and galaxy-galaxy lensing models according to,
\begin{equation}
    \xi_{+/-}^{ij}(\theta) = (1+m_i) \, (1+m_j) \, \xi^{ij}_{+/-,\text{true}}(\theta)
\end{equation}
and
\begin{equation}
    \gamma^{ij}_{t}(\theta) = (1+m_j) \, \gamma^{ij}_{t,\text{true}}(\theta) .
\end{equation}
One multiplicative shear bias parameter is included for each tomographic source bin.  We again replicate the Gaussian priors in the shear bias parameters used by the different weak lensing survey collaborations, as listed in Table~\ref{tab:priors} and referenced below.  We note that our mocks do not contain the effect of blending, which can create a coupling between the shear bias and redshift distributions \citep{2022MNRAS.509.3371M}.

\textit{Lens magnification.}  We model magnification effects in the selection of the lens galaxy sample following \cite{2023MNRAS.523.3649E}, where a proportionality is assumed between the density contrast due to lens magnification and the convergence field,
\begin{equation}
    \delta_{\mu,l}^i = 2 \, (\alpha_l^i-1) \, \kappa_l^i .
\end{equation}
The lens magnification parameters $\alpha_l^i$ can be accurately measured for each lens bin $i$ from the Buzzard mocks following \cite{2024OJAp....7E..57L}, and we fix these coefficients during our fits.  We note that there can be systematic errors associated with redshift evolution of these coefficients in the context of real data samples \citep{2024MNRAS.527.1760W}, although we neglect these issues in our current simulation-based analysis.

We summarise the priors assumed in our analysis in Table~\ref{tab:priors}, which are based on \cite{2020PASJ...72...16H} and \cite{2023OJAp....6E..36D}.  We note that our parameter set excludes the effects of massive neutrinos, as we do not account for that in our baseline simulation, as well as parameters associated with the intrinsic alignment of galaxies, which are not infused into the mocks we adopt for this study.  The choice of priors therefore does not exactly match our approach for the real data analysis \citep{AnnaDESI}, in which we will marginalise over intrinsic alignment models following \cite{2023OJAp....6E..36D}, but nonetheless allows us to create realistic end-to-end pipeline tests.

\begin{table}[t]
    \centering
    {\renewcommand{\arraystretch}{1.2}
    \begin{tabular}{cccc}
        \hline
        \multicolumn{4}{l}{\textbf{Common parameters for all lensing surveys}} \\
        \hline
        \multicolumn{4}{l}{\textbf{Cosmological parameters:}} \\
        $\Omega_m$ & \multicolumn{3}{l}{flat (0.1,0.9)} \\
        $\Omega_b$ & \multicolumn{3}{l}{flat (0.03,0.07)} \\
        $H_0$ & \multicolumn{3}{l}{flat (55,91) km s$^{-1}$ Mpc$^{-1}$ } \\
        $A_s$ & \multicolumn{3}{l}{flat (0.5,5.0) $\times 10^{-9}$} \\
        $n_s$ & \multicolumn{3}{l}{flat (0.87,1.07)} \\
        \hline
        \multicolumn{4}{l}{\textbf{Galaxy bias:}} \\
        $b_1^i$  & \multicolumn{2}{l}{flat (0.8,3.0)} & $i=1,2,...,5$  \\
        \hline
        \multicolumn{4}{l}{\textbf{Lens magnification:}} \\
        $\alpha_l^1$ & \multicolumn{3}{l}{Fixed to 0.91} \\
        $\alpha_l^2$ & \multicolumn{3}{l}{Fixed to 1.58} \\
        $\alpha_l^3$ & \multicolumn{3}{l}{Fixed to 2.02} \\
        $\alpha_l^4$ & \multicolumn{3}{l}{Fixed to 2.58} \\
        $\alpha_l^5$ & \multicolumn{3}{l}{Fixed to 2.26} \\
        \hline
        & \textbf{DES} & \textbf{KiDS} & \textbf{HSC} \\
        \hline
        \multicolumn{4}{l}{\textbf{Photometric redshift uncertainties:}} \\
        $\Delta z_s^1 \times 100$ & $\mathcal{N}(0,1.8)$ & $\mathcal{N}(0,1.1)$ & $\mathcal{N}(0,3.74)$ \\
        $\Delta z_s^2 \times 100$ & $\mathcal{N}(0,1.5)$ & $\mathcal{N}(0,1.1)$ & $\mathcal{N}(0,1.24)$ \\
        $\Delta z_s^3 \times 100$ & $\mathcal{N}(0,1.1)$ & $\mathcal{N}(0,1.2)$ & $\mathcal{N}(0,3.26)$ \\
        $\Delta z_s^4 \times 100$ & $\mathcal{N}(0,1.7)$ & $\mathcal{N}(0,0.9)$ & $\mathcal{N}(0,3.43)$ \\
        $\Delta z_s^5 \times 100$ &  & $\mathcal{N}(0.0,1.0)$ &  \\
        \hline
        \multicolumn{4}{l}{\textbf{Multiplicative shear bias:}} \\
        $m_1 \times 100$ & $\mathcal{N}(0,0.9)$ & $\mathcal{N}(0.0,1.9)$ & $\mathcal{N}(0.0,1.0)$ \\
        $m_2 \times 100$ & $\mathcal{N}(0,0.8)$ & $\mathcal{N}(0.0,2.0)$ & $\mathcal{N}(0.0,1.0)$ \\
        $m_3 \times 100$ & $\mathcal{N}(0,0.8)$ & $\mathcal{N}(0.0,1.7)$ & $\mathcal{N}(0.0,1.0)$ \\
        $m_4 \times 100$ & $\mathcal{N}(0,0.8)$ & $\mathcal{N}(0.0,1.2)$ & $\mathcal{N}(0.0,1.0)$ \\
        $m_5 \times 100$ &  & $\mathcal{N}(0.0,1.0)$ &  \\
        \hline
    \end{tabular}
    }
    \caption{The parameters and priors assumed for the model fits to the mock datasets.  The upper section of the table lists the parameters common to the analysis of all lensing surveys: the cosmological parameters, galaxy bias parameters and lensing magnification coefficients.  We use different photometric redshift uncertainties and multiplicative shear bias for each lensing survey, assuming a Gaussian distribution denoted by $\mathcal{N}(\mu,\sigma)$, as listed in the lower section of the table.}
    \label{tab:priors}
\end{table}

\begin{table*}[t]
    \centering
    {\renewcommand{\arraystretch}{1.4}
\begin{tabularx}{\textwidth}{c|*{4}{>{\centering\arraybackslash}X}|*{4}{>{\centering\arraybackslash}X}}
    \multicolumn{8}{c}{\textbf{DESI $\times$ HSC-Y1}} \\ \hline\hline
    & \multicolumn{4}{c|}{$R_\text{clus} = 7 \, h^{-1}$ Mpc, $R_\text{ggl} = 6 \, h^{-1}$ Mpc} & \multicolumn{4}{c}{$R_\text{clus} = 7 \, h^{-1}$ Mpc, $R_\text{ggl} = 10 \, h^{-1}$ Mpc} \\ \cline{2-9}
    & DoF & $\langle \chi^2 \rangle$ & $\chi^2 (\chi_\text{nosys}^2)$ & bias in $\Omega_m$-$S_8$ & DoF & $\langle \chi^2 \rangle$ & $\chi^2 (\chi_\text{nosys}^2)$ & bias in $\Omega_m$-$S_8$ \\ \hline
    $\xi_\pm$ & 158 & 177.4 & 188.2 (11.7) & 0.02$\sigma$ & -- & -- & -- & -- \\
    $\gamma_t + w_p$ & 88 & 113.2 & 222.6 (152.5) & 1.09$\sigma$ & 70 & 101.1 & 209.1 (148.6) & 1.06$\sigma$ \\
    $3\times 2$-pt & 258 & 300.1 & 433.6 (392.3) & 1.04$\sigma$ & 240 & 287.0 & 413.5 (376.3) & 0.99$\sigma$ \\ \hline \hline
    \multicolumn{8}{c}{}\\
    \multicolumn{8}{c}{\textbf{DESI $\times$ KiDS-1000}} \\ \hline\hline
    & \multicolumn{4}{c|}{$R_\text{clus} = 7 \, h^{-1}$ Mpc, $R_\text{ggl} = 6 \, h^{-1}$ Mpc} & \multicolumn{4}{c}{$R_\text{clus} = 7 \, h^{-1}$ Mpc, $R_\text{ggl} = 10 \, h^{-1}$ Mpc} \\ \cline{2-9}
    & DoF & $\langle \chi^2 \rangle$ & $\chi^2 (\chi_\text{nosys}^2)$ & bias in $\Omega_m$-$S_8$ & DoF & $\langle \chi^2 \rangle$ & $\chi^2 (\chi_\text{nosys}^2)$ & bias in $\Omega_m$-$S_8$ \\ \hline
    $\xi_\pm$ & 210 & 238.9 & 232.1 (8.1) & 0.55$\sigma$ & -- & -- & -- & -- \\
    $\gamma_t + w_p$ & 108 & 108.9 & 101.8 (44.9) & 0.73$\sigma$ & 88 & 94.0 & 88.8 (43.5) & 0.62$\sigma$ \\
    $3\times 2$-pt & 333 & 352.4 & 338.9 (295.4) & 0.85$\sigma$ & 313 & 334.2 & 326.6 (285.7) & 0.80$\sigma$ \\ \hline \hline
    \multicolumn{8}{c}{}\\
    \multicolumn{8}{c}{\textbf{DESI $\times$ DES-Y3}} \\ \hline\hline
    & \multicolumn{4}{c|}{$R_\text{clus} = 7 \, h^{-1}$ Mpc, $R_\text{ggl} = 6 \, h^{-1}$ Mpc} & \multicolumn{4}{c}{$R_\text{clus} = 7 \, h^{-1}$ Mpc, $R_\text{ggl} = 10 \, h^{-1}$ Mpc} \\ \cline{2-9}
    & DoF & $\langle \chi^2 \rangle$ & $\chi^2 (\chi_\text{nosys}^2)$ & bias in $\Omega_m$-$S_8$ & DoF & $\langle \chi^2 \rangle$ & $\chi^2 (\chi_\text{nosys}^2)$ & bias in $\Omega_m$-$S_8$ \\ \hline
    $\xi_\pm$ & 215 & 205.6 & 212.6 (8.1) & 0.53$\sigma$ & -- & -- & -- & -- \\
    $\gamma_t + w_p$ & 88  & 88.0 & 124.0 (47.7) & 0.82$\sigma$ & 70 & 75.7 & 108.6 (42.0) & 0.48$\sigma$ \\
    $3\times 2$-pt & 315 & 296.6 & 347.0 (297.2) & 0.86$\sigma$ & 297 & 284.4 & 326.4 (266.7) & 0.95$\sigma$ \\ \hline \hline
\end{tabularx}
    }
    \caption{Results summary for the configurations tested in this work.  We consider cases varying the mock lensing dataset (HSC-Y1, KiDS-1000 and DES-Y3) and the scale cuts applied when fitting the GGL and clustering measurements, $R_\text{ggl}$ and $R_\text{clus}$.  We also consider three combinations of correlation functions: cosmic shear ($\xi_\pm$), $2\times 2$-pt functions ($\gamma_t+w_p$), and the $3\times 2$-pt functions ($\xi_\pm+\gamma_t+w_p$), in all cases assuming the covariance for a single mock realisation.  The four columns within each panel show respectively the number of degrees of freedom (DoF), the average value $\langle \chi^2 \rangle$ of fits to individual mock realisations, the best-fit $\chi^2$ value for the mock mean data vector (and best-fit $\chi^2_\text{nosys}$ value for the noise-free mock mean) multiplied by a factor of 8 to account for the number of realisations, and the parameter bias between the best-fitting values of $\{ \Omega_m, S_8 \}$ and the fiducial cosmology in the $\Omega_m$-$S_8$ plane, as defined in Sec.~\ref{sec:pte}.  The choice of $R_\text{ggl}$ and $R_\text{clus}$ has no impact on the cosmic shear fits.}
    \label{tab:chi2_fits}
\end{table*}

\begin{table*}[t]
    \centering
    {\renewcommand{\arraystretch}{1.4}
    \begin{tabular}{c|c|c|c|c|c|c|c|c|c|c|c|c}
     \hline\hline
     \textbf{Survey:} & \multicolumn{4}{c|}{HSC-Y1} & \multicolumn{4}{c|}{KiDS-1000} & \multicolumn{4}{c}{DES-Y3} \\ \cline{2-13}
     \textbf{$\Omega_m$-$S_8$ bias:} & 0.3$\sigma$ & 0.5$\sigma$ & 1$\sigma$ & 2$\sigma$ & 0.3$\sigma$ & 0.5$\sigma$ & 1$\sigma$ & 2$\sigma$ & 0.3$\sigma$ & 0.5$\sigma$ & 1$\sigma$ & 2$\sigma$ \\ \hline
     $\xi_\pm$ & 0.34 & 0.12 & <0.01 & <0.01 & 0.56 & 0.23 & <0.01 & <0.01 & 0.56 & 0.28 & 0.02 & <0.01 \\ \hline
     $\gamma_t+w_p$ & 0.83 & 0.63 & 0.16 & <0.01 & 0.41 & 0.21 & 0.03 & <0.01 & 0.49 & 0.24 & 0.03 & <0.01 \\ \hline
     $3\times 2$-pt & 0.86 & 0.70 & 0.28 & <0.01 & 0.22 & 0.07 & <0.01 & <0.01 & 0.75 & 0.42 & 0.04 & <0.01 \\ \hline\hline
    \end{tabular}
    }
    \caption{Values of the Probability-to-Exceed (PTE) statistic described in Sec.~\ref{sec:pte}, which describes the probability that parameter fits to the mock mean data vector lie outside those obtained from fits applied to the fiducial model, using the same covariance and parameter set.  Hence, the PTE statistic includes potential parameter projection (prior volume) effects.  We show PTE values for a cosmological parameter bias of 0.3$\sigma$, 0.5$\sigma$, 1$\sigma$ and 2$\sigma$.}
    \label{tab:PTE}
\end{table*}

\begin{figure*}
\centering
\includegraphics[width=\textwidth]{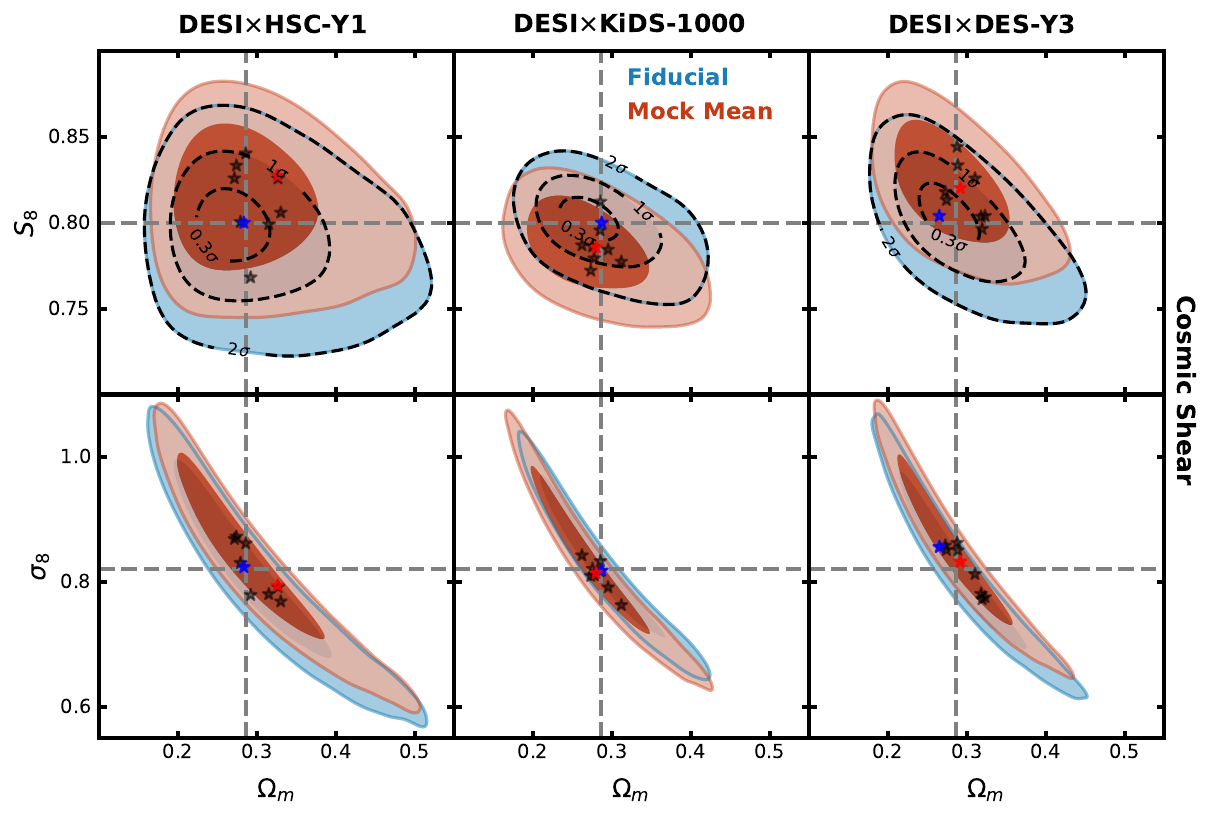}
\caption{1$\sigma$ and 2$\sigma$ confidence contours derived from cosmic shear cosmological fits, represented in the projections of $\Omega_m$-$S_8$ (top row) and $\Omega_m$-$\sigma_8$ (bottom row).  The blue contours represent the fits to a synthetic data vector generated using the fiducial cosmology of the Buzzard simulation suite, and the red contours correspond to fits to the mock mean data vector. In the case of the contours for $\Omega_m$-$S_8$, we also show the 0.3$\sigma$ confidence contours for the fits to the fiducial model data vector. These black, dashed contours are used for calculating the PTE values discussed in Sec.~\ref{sec:pte} and Table~\ref{tab:PTE}. Additionally, we include the best-fit values obtained from $\chi^2$ minimization for the fiducial model data vector (blue stars), the mock mean data vector (red stars) and for each of the individual mock realisations (black stars). The intersection of the dashed grey lines represents the fiducial parameter values of the Buzzard simulation suite.}
\label{fig:cosmic_shear_contours}
\vspace{0.5cm}
\end{figure*}

\begin{figure*}
\centering
\includegraphics[width=\textwidth]{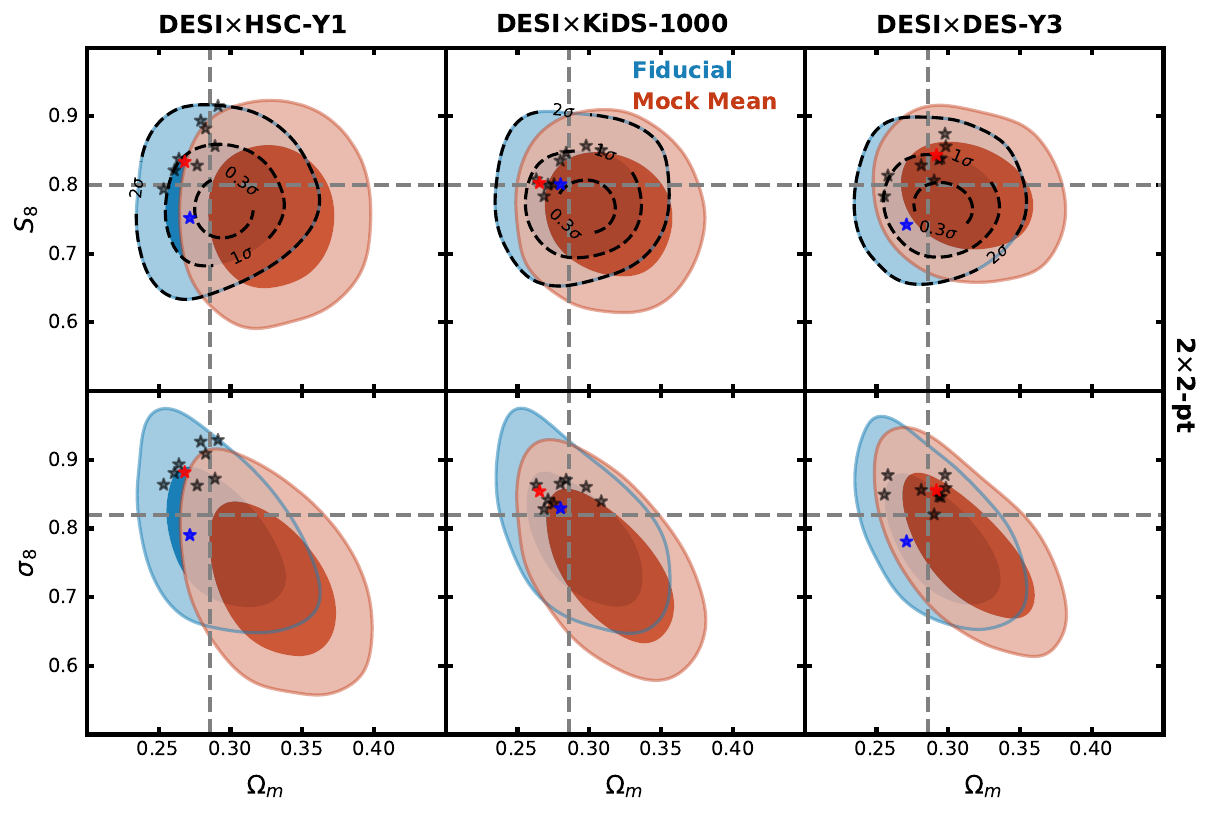}
\caption{1$\sigma$ and 2$\sigma$ confidence contours derived from cosmological fits to a $\gamma_t(\theta)+w_p(R)$ data vector representing the fiducial model (blue contours) and the mock mean (red contours).  These results use scale cuts $R_\text{clus} = 7 \, h^{-1}$ Mpc and $R_\text{ggl} = 10 \, h^{-1}$ Mpc and are displayed in the same format as Fig.~\ref{fig:cosmic_shear_contours}.}
\label{fig:2x2pt_contours}
\vspace{0.5cm}
\end{figure*}

\begin{figure*}
\centering
\includegraphics[width=\textwidth]{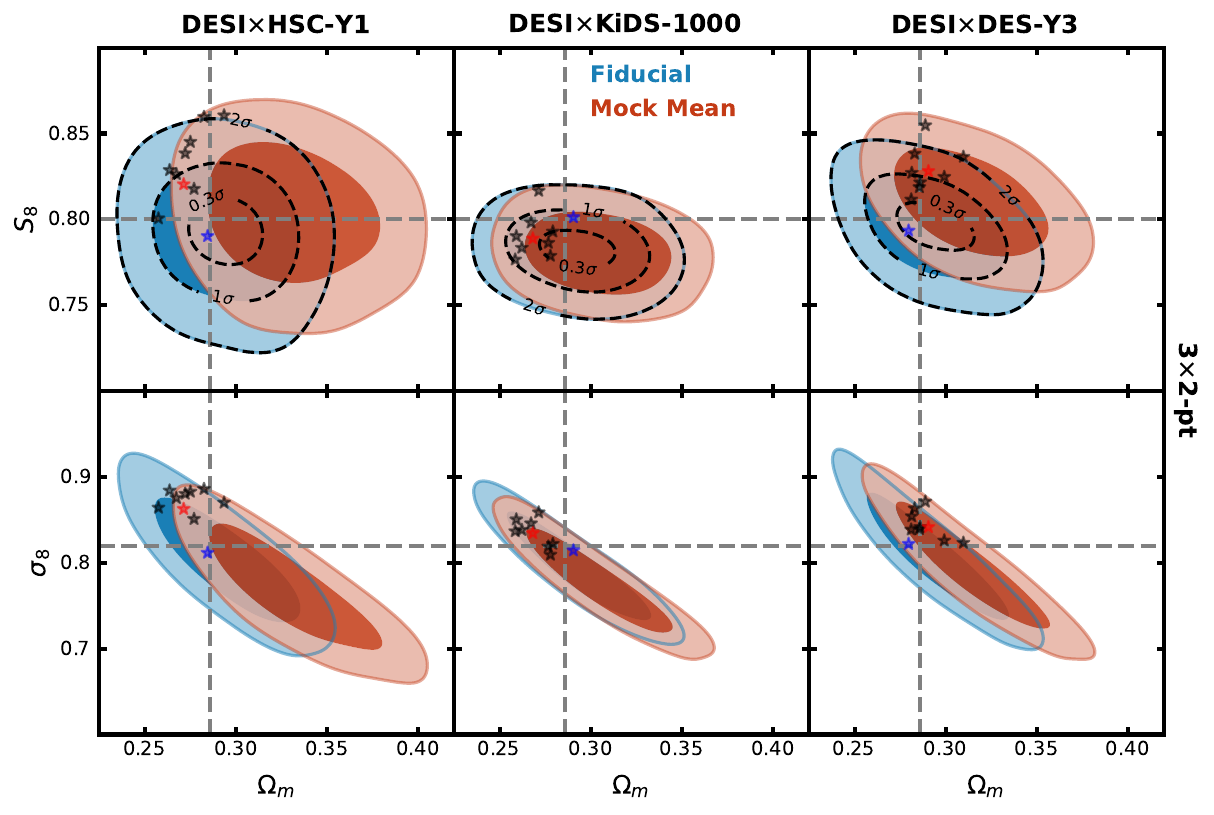}
\caption{1$\sigma$ and 2$\sigma$ confidence contours derived from cosmological fits to a $3 \times 2$-pt data vector representing the fiducial model (blue contours) and the mock mean (red contours).  These results use scale cuts $R_\text{clus} = 7 \, h^{-1}$ Mpc and $R_\text{ggl} = 10 \, h^{-1}$ Mpc, and the same cosmic shear scale cuts as each survey collaboration, and are displayed in the same format as Fig.~\ref{fig:cosmic_shear_contours}.}
\label{fig:3x2pt_contours}
\vspace{0.5cm}
\end{figure*}

\section{Cosmological parameter recovery}
\label{sec:cosmo}

In this section we perform tests of our recovery of the input cosmological parameters using our simulated measurements, covariance and theoretical models.  We perform these tests independently for each simulated weak lensing survey.  For our fitting platform, we use {\tt CosmoMC} \citep{2013PhRvD..87j3529L}.

For our fiducial analysis we fit the mean data vector obtained from averaging together the 8 different Buzzard mock simulations of each lensing survey, where each individual realisation is representative of the DESI-Y1 dataset including overlapping and non-overlapping regions corresponding to cosmic shear, GGL and clustering, as described in Sec.~\ref{sec:datavectors}. In these fits we do not re-scale the analytical covariance matrix by the number of mock realisations, rather we use the covariance corresponding to a single realisation, since this yields statistical errors comparable to the DESI-Y1 analysis.  We exclude the third lens bin from our analysis described below, due to the fact that two light cone simulations transition within this lens bin, as noted in Sec.~\ref{sec:simulations}, leading to spurious effects on the correlation functions.

\subsection{Scale cuts}

Cosmological analyses typically apply scale cuts to exclude separation ranges where the measurements or models might be afflicted by systematic errors.  A primary concern in our case is that the linear galaxy bias model we adopt in the clustering and galaxy-galaxy lensing models is not applicable on scales smaller than several $h^{-1}$ Mpc \citep[e.g.,][]{2021A&A...646A.129J, 2021arXiv210513548K, 2022PhRvD.105l3520D}.  Hence, we consider restricting our fits to ranges of projected separation $R > R_\text{ggl}$ for the average tangential shear, and $R > R_\text{clus}$ for the projected correlation function, converting projected to angular separations using $\theta = R/\chi(\overline{z_l})$, where $\overline{z_l}$ is the mean redshift of each lens bin.  Hence, we allow for different scale cuts for these two probes, given their differing bias dependence and signal-to-noise ratios, and the fact that $\gamma_t(\theta)$ represents a ``non-local'' signal depending on smaller scales \citep{2020MNRAS.491.5498M}.  In the case of $w_p(R)$, we also apply a maximum scale cut which removes any effects from the domain decomposition used when constructing the galaxy catalogue, which we have conservatively estimated as half the projected separation of the average angular size of the $n_\text{side}=8$ pixels.

For the cosmic shear correlation functions, we apply the same scale cuts as used by each weak lensing survey collaboration.  For KiDS-1000, we analyse $\xi_+(0.5 < \theta < 300 \; \mathrm{arcmin})$ and $\xi_-(4 < \theta < 300 \; \mathrm{arcmin})$ \citep{2021A&A...645A.104A}, and for HSC-Y1 we analyse $\xi_+(7.08 < \theta < 56.2 \; \mathrm{arcmin})$ and $\xi_-(28.2 < \theta < 178 \; \mathrm{arcmin})$ \citep{2020PASJ...72...16H}.  For DES-Y3 we apply the same scale cuts as optimised for $\Lambda$CDM fits by \cite{2021arXiv210513548K}, in which the overall separation range $2.5 < \theta < 250 \; \mathrm{arcmin}$ is restricted for each separate pair of tomographic source bins, according to a $\Delta \chi^2$ threshold determined by comparing a baseline model to a ``contaminated'' model containing reasonable physical variations.

Our fiducial choices of the scale cuts are $R_\text{clus} = 7 \, h^{-1}$ Mpc and $R_\text{ggl} = 10 \, h^{-1}$ Mpc, as described below, although we also consider a case with $R_\text{ggl} = 6 \, h^{-1}$ Mpc.  We refer to the upcoming study by \cite{NimasDESI} for the further validation of scale cuts in the context of the full DESI-Y1 $3 \times 2$-pt analysis, in which we incorporate noise-free tests and point-mass marginalisation.

\subsection{Tests with fixed cosmology}
\label{sec:fixedcosmo}

First, we tested whether our theoretical models were able to produce a good fit to the data whilst setting the cosmological parameters to the fiducial values of the Buzzard simulation suite.  Varying the systematic nuisance parameters using the priors assumed in Table \ref{tab:priors} leads to minimum reduced $\chi^2$ values of 0.96, 1.07 and 1.25 for DES-Y3, KiDS-1000 and HSC-Y1, respectively, where we average over the results of fits performed to the 8 independent realisations.  Here, the reduced $\chi^2$ is defined as the best-fit $\chi^2$ divided by the associated number of degrees of freedom, which we estimate as $\text{DoF}=n_d-n_p$, where $n_\text{d}$ is the number of data points and $n_\text{p}$ is the number of fitted parameters\footnote{We note that this estimate is only an approximation of the effective number of degrees of freedom, as discussed by \cite{2021A&A...646A.129J}.}.  If we vary the lens magnification parameter values in addition to the other nuisance parameters, we find that we recover values consistent with the fixed inputs listed in Table \ref{tab:priors}.

\subsection{Parameter recovery statistics}
\label{sec:pte}

In order to quantify the statistical recovery of the fiducial cosmological parameters, we focus on the most important 2D projections of the parameter space -- that is, $\{ \Omega_m, \sigma_8 \}$ and its alternative parameterisation, $\{ \Omega_m, S_8 \}$.  We use two metrics to quantify the agreement between the 2D parameter contours and the fiducial point.

First, we define the \textit{parameter bias}, $N$-$\sigma$, as the \textit{marginalised confidence interval of the fit to the mock mean data vector, using the unscaled covariance, which intersects the fiducial point} ($\Omega_m = 0.286$, $\sigma_8 = 0.82$).  We note two limitations of this metric.  First, the limited number of mocks creates a statistical noise in this estimate, such that the measured parameter bias is not zero in the absence of systematic errors (we can estimate the noise in the estimate from 8 realisations as $\frac{1}{\sqrt{8}} \approx 0.4\sigma$).  Second, when comparing likelihood contours marginalised over many other cosmological and nuisance parameters, we should be cautious of ``parameter projection effects'' \citep[as discussed for example by][]{2021arXiv210513548K}, which can lead to apparent shifts in projected likelihood contours induced by the multi-dimensional prior volume.

In order to mitigate against these two issues we follow the approach of \cite{2022PhRvD.105l3520D}, leading us to define a second deviation metric.  As well as fitting cosmological models to the mock mean data vectors $\boldsymbol{d}_\text{mock}$ (which contain noise and potential systematics), we also run these fits on a fiducial model data vector $\boldsymbol{d}_\text{fid}$, using the same parameterisation and covariance (which should naturally result in a low $\chi^2 \approx 0$ since there is no noise in the model vector).  Any prior volume effects will manifest in an apparent offset of the resulting 2D projected parameter contours of the fiducial model fit relative to the fiducial point, hence the comparison of the fits to the mock mean and fiducial model vectors allows us to quantify parameter systematics.

We quantify the deviation between these fits using the \textit{Probability-to-Exceed} (PTE) statistic \citep{2022PhRvD.105l3520D} defined as,
\begin{equation}
\begin{split}
    \text{PTE} & = 1 - \int_{\mathcal{R}} P(S_8,\Omega_m|\boldsymbol{d}_\text{mock};\frac{C}{M}) \, d^2\theta \\
    & \approx 1 - \frac{\int_{\mathcal{R}} P(S_8,\Omega_m|\boldsymbol{d}_\text{mock};C)^M \, d^2\theta}{\int_{\mathrm{all} \, \theta} P(S_8,\Omega_m|\boldsymbol{d}_\text{mock};C)^M \, d^2\theta},
\end{split}
\label{eq:PTE}
\end{equation}
where $\theta = \left( S_8, \Omega_m \right)$ is the projected parameter space, and $\mathcal{R}$ defines a region in this parameter space bounded by the $N$-$\sigma$ contour of $P(S_8,\Omega_m|\boldsymbol{d}_\text{fid};C)$ enclosing a total probability $\pi_{\text{thres}}$.  Also in Eq.~\ref{eq:PTE}, $C$ is the covariance matrix for a single realisation, $M$ is the number of mock realisations, and the second line follows from an assumption that the prior in the 2D parameter space is flat over the region in question \citep{2022PhRvD.105l3520D}.  Hence, Eq.~\ref{eq:PTE} represents an integral over $S_8$ and $\Omega_m$ of the 2D posterior distribution of the fit to the mock mean data vector, across an interval defined by the $N$-$\sigma$ confidence region of the 2D posterior distribution of the fit to the fiducial model data vector.

In summary, the PTE represents the \textit{probability that the analysis performed on the mock mean data vector results in a parameter fit outside the $N$-$\sigma$ confidence interval of the same analysis performed on the fiducial model data vector using the unscaled covariance.}  A low value of PTE (close to 0) indicates that the fits to the fiducial model and mock mean data vectors are in a good agreement, such that the parameters derived from the mock mean data are unlikely to stray outside those obtained from the fiducial model data.  On the other hand, a high PTE value (close to 1) suggests that the fit to the mock mean data does not agree well with the fiducial model data fit.

In this work, we test four thresholds: a strict threshold of no greater than $0.3\sigma$ parameter deviation (referenced to a Gaussian distribution), which corresponds to $\pi_{\text{thres}} \approx 0.23$ \citep{2022PhRvD.105l3520D}, moderate thresholds of $0.5\sigma$ and $1\sigma$, where $\pi_{\text{thres}} \approx 0.38$ and $0.68$, respectively, and a $2\sigma$ threshold with $\pi_{\text{thres}} \approx 0.95$. These threshold values lead to integrating the renormalized posterior $P(S_8,\Omega_m|\boldsymbol{d}_\text{mock};C)^M$ around the corresponding $N$-$\sigma$ region defined by $P(S_8,\Omega_m|\boldsymbol{d}_\text{fid};C)$.  We calculated PTE values for all the fits described in the following sections, as displayed in Table \ref{tab:PTE}.

\subsection{Cosmic shear-only fits}

We initially fit the cosmic shear measurements only, and in Fig.~\ref{fig:cosmic_shear_contours} we show the marginalised confidence regions for $\Omega_m$ and $\sigma_8$ for the different simulated weak lensing surveys as the red contours.  We recover the fiducial values of these parameters within the statistical error margins in all the fits.  For the cases of (HSC-Y1, KiDS-1000, DES-Y3) we find that the averaged best-fitting $\chi^2$ of individual mock realisations is $(177.4, 238.9, 205.6)$ for $\text{DoF} = (158, 210, 215)$, and the parameter bias values are $(0.02, 0.55, 0.53) \sigma$.  These results are all displayed in Table \ref{tab:chi2_fits}, in which we also provide the minimum $\chi^2$ values for fits to measurements without shape noise and lens magnification applied.  As we noted in Sec.~\ref{sec:pte}, the parameter bias values are subject to noise, so we also consider the Probability-to-Exceed values of the fits relative to those of a fiducial model data vector.  We find that the PTE for a 0.3$\sigma$ bias is between 11\% and 56\% depending on the case, and  drops below 2\% for a 1$\sigma$ bias.  Full results are listed in Table \ref{tab:PTE}.

Fig.~\ref{fig:cosmic_shear_contours} also displays the fits to the fiducial model data vectors, which are used in the PTE calculation, as the blue contours.  Additionally, for each case we show the best-fitting parameter values for the mock mean data (red star), the individual mock realisations (black stars) and the fiducial model data vector (blue star).  We find that all the fits to the individual mocks and mean mock are consistent with the red contours.  The dashed regions within the blue contours represent the intervals used for the calculation of the PTE values.  However, we note that the PTE calculation effectively uses the rescaled covariance matrix (as expressed in Eq.~\ref{eq:PTE}) such that the values themselves cannot be easily inferred from Fig.~\ref{fig:cosmic_shear_contours}.

\subsection{Fits to GGL and clustering}

Next, we consider fitting to the galaxy-galaxy lensing and clustering measurements, $\gamma_t+w_p$ (or ``$2 \times 2$-pt'' data).  Since these statistics both involve modelling galaxy bias, we apply scale cuts where a linear bias model might not be valid at the level of precision of our measurement, removing separation ranges $R < R_\text{ggl}$ (for GGL) and $R < R_\text{clus}$ (for clustering).  We again performed fits to the mean data vectors derived from 8 Buzzard mock simulations, using a single-realisation covariance.  The parameter fits are shown in Fig.~\ref{fig:2x2pt_contours}, and a summary of the results for the various configurations tested can be found in Tables \ref{tab:chi2_fits} and \ref{tab:PTE}.

Adopting our fiducial scale cuts $R_\text{clus} = 7 \, h^{-1}$ Mpc and $R_\text{ggl} = 10 \, h^{-1}$ Mpc, we find that the averaged best-fitting $\chi^2$ values of the $2 \times 2$-pt individual mock realisations for (HSC-Y1, KiDS-1000, DES-Y3) are $(101.1, 94.0, 75.7)$ for $\text{DoF} = (70, 88, 70)$, and the parameter bias values are $(1.06, 0.62, 0.48) \sigma$.  These biases are somewhat higher than those obtained for cosmic-shear-only fits, and we find PTE values $(81\%, 41\%, 49\%)$ for a 0.3$\sigma$ threshold and $(22\%, 3\%, 3\%)$ for a 1$\sigma$ threshold.  The performance of the HSC-Y1 fits is somewhat worse than the other two simulated surveys; we speculate that this effect is caused by the impact of the survey boundaries of the small HSC-Y1 regions on the $\gamma_t(\theta)$ covariance, as discussed and quantified by \cite{2024MNRAS.533..589Y}, rather than by our modelling framework.

For these scale cuts, we used the parameter differences tension estimator $Q_\text{UDM}$ from \cite{2020PhRvD.101j3527R} to determine that the significance of the tension between $\xi_\pm$ and $\gamma_t+w_p$ is below 0.3$\sigma$.  Fig.~\ref{fig:2x2pt_contours} suggests the presence of some projection effects in our $2 \times 2$-pt fits in the 2D contours for $S_8$ and $\Omega_m$, since the best-fitting values (represented by stars) can sometimes be located outside the 1$\sigma$ contours for the mock mean data vector case, especially for the case of HSC-Y1.

We conclude from these investigations that our modelling of the cosmic shear within the simulation is somewhat better than our modelling of $\gamma_t+w_p$, although the latter is still adequate to ensure that systematic errors are sub-dominant to statistical errors.  We found that adopting the less conservative GGL scale cut $R_\text{ggl} = 6 \, h^{-1}$ Mpc produced slightly higher inconsistencies between $\xi_\pm$ and $\gamma_t+w_p$ and parameter biases (especially for DES-Y3).  We also summarise our results for this configuration in Table \ref{tab:chi2_fits}.  We found that adopting a more conservative clustering scale cut $R_\text{clus} = 10 \, h^{-1}$ Mpc did not significantly affect parameter recovery.

\subsection{Fits to the full data vector}

Finally, we extended our fits to the full $3 \times 2$-pt correlation functions, jointly fitting $\xi_{\pm}$, $\gamma_t$ and $w_p$.   Our marginalised results for $\Omega_m$ and $\sigma_8$, along with the parameter combination $S_8$, are shown in Fig.~\ref{fig:3x2pt_contours}, and a summary of the results is given in Tables \ref{tab:chi2_fits} and \ref{tab:PTE}.

Adopting our fiducial scale cuts as above, we find that the averaged best-fitting $\chi^2$ values of the $3 \times 2$-pt individual mock realisations for (HSC-Y1, KiDS-1000, DES-Y3) are $(287.0, 334.2, 284.4)$ for $\text{DoF} = (240, 313, 297)$, and the parameter bias values are $(0.99, 0.80, 0.95) \sigma$, where we continue to note that these values contain statistical noise and parameter projection effects, in addition to representing any systematics.  We find that the Probability-to-Exceed is $(82\%, 22\%, 75\%)$ for a 0.3$\sigma$ bias and $(25\%, 1\%, 4\%)$ for a 1$\sigma$ bias.  These values are comparable to those reported by \cite{2022PhRvD.105l3520D} for their end-to-end analysis of DES-Y3.  We conclude from these PTE values, and the comparable performance to previous work, that our fiducial $3 \times 2$-pt fitting configuration produces an acceptable recovery of the underlying cosmological parameters at the level of precision appropriate for the DESI-Y1 dataset.  We show the best-fit 3$\times$2-pt correlation functions for all the cases in Appendix \ref{sec:fixed_cosmo_fits}.

We close this section by commenting that these results are naturally contingent on the extent to which the $N$-body simulations we utilise fully capture the cosmological, astrophysical and observational effects that are present in the real data samples.  Given that simulations may not perfectly encode all potential systematic effects, in future work we will re-visit the scale cut choices using other techniques such as ``analytical'' noise-free tests \citep{2021arXiv210513548K}.

\section{Conclusions}
\label{sec:conc}

In this paper we have presented an end-to-end validation of cosmological parameter recovery from $3 \times 2$-pt correlation functions, based on realistic simulations of a combined-probe experiment featuring the DESI-Y1 large-scale structure dataset and partially-overlapping KiDS, DES and HSC weak lensing datasets.  Our mock catalogues, derived from the Buzzard $N$-body simulations, include realistic photometric redshift errors, source weights, shape noise, magnification, multiplicative shape bias and lens clustering.  We use the projected correlation function $w_p(R)$ to quantify the galaxy clustering, reflecting the availability of spectroscopic redshifts.

We have provided a detailed description of the determination of the analytical covariance including, for the first time, the cross-covariance between the angular cosmic shear and galaxy-galaxy lensing statistics and the projected correlation function.  We validate the analytical covariance through comparison with the ensemble of simulations, and external codes.

We illustrate how the {\tt CosmoMC} cosmological fitting platform can use these simulated data vectors and covariances to recover values of the cosmological parameters with systematic errors sub-dominant to the statistical errors, and test the appropriate choices of fitting ranges on the assumption of linear galaxy bias.  However, we point out that model extensions such as point-mass marginalization \citep{2020MNRAS.491.5498M} and non-linear galaxy bias \citep{2022PhRvD.105l3520D} could further improve the cosmological parameter recovery or enable smaller separation scales to be accessed by fitting.

This paper lays the groundwork for an upcoming combined-probe analysis with the DESI-Y1, KiDS, DES and HSC datasets, which will provide new, accurate constraints on key cosmological parameters.  Future work will present scale cut tests using analytical model variations and shear ratio validations using the survey datasets \citep{NimasDESI}, together with our final $3 \times 2$-pt cosmological analyses of the DESI-Y1 data \citep{AnnaDESI}.  We also plan to extend these studies to joint analyses of lensing and three-dimensional clustering information, including redshift-space distortions.  Such cosmological analyses, and their validation, will continue to be significant as lensing and large-scale structure datasets grow, enabling ever-more stringent tests of the consistency of our cosmological model.

\section*{Acknowledgements}

We thank Peter Taylor, Jamie McCullough and two anonymous referees for useful comments on a draft of this paper.

CB acknowledges financial support received through Australian Research Council Discovery Project DP220101609. CGQ acknowledges support provided by NASA through the NASA Hubble Fellowship grant HST-HF2-51554.001-A awarded by the Space Telescope Science Institute, which is operated by the Association of Universities for Research in Astronomy, Inc., for NASA, under contract NAS5-26555.

This material is based upon work supported by the U.S. Department of Energy (DOE), Office of Science, Office of High-Energy Physics, under Contract No. DE–AC02–05CH11231, and by the National Energy Research Scientific Computing Center, a DOE Office of Science User Facility under the same contract. Additional support for DESI was provided by the U.S. National Science Foundation (NSF), Division of Astronomical Sciences under Contract No. AST-0950945 to the NSF’s National Optical-Infrared Astronomy Research Laboratory; the Science and Technology Facilities Council of the United Kingdom; the Gordon and Betty Moore Foundation; the Heising-Simons Foundation; the French Alternative Energies and Atomic Energy Commission (CEA); the National Council of Humanities, Science and Technology of Mexico (CONAHCYT); the Ministry of Science, Innovation and Universities of Spain (MICIU/AEI/10.13039/501100011033), and by the DESI Member Institutions: \url{https://www.desi.lbl.gov/collaborating-institutions}. Any opinions, findings, and conclusions or recommendations expressed in this material are those of the author(s) and do not necessarily reflect the views of the U. S. National Science Foundation, the U. S. Department of Energy, or any of the listed funding agencies.

The authors are honored to be permitted to conduct scientific research on Iolkam Du’ag (Kitt Peak), a mountain with particular significance to the Tohono O’odham Nation.

\section*{Data availability}

Data points for all the figures are available at \url{https://doi.org/10.5281/zenodo.14248798}.

\bibliographystyle{mnras}
\bibliography{desi_3x2pt_mockchallenge}

\section*{Affiliations}
\scriptsize
\noindent
$^{1}$ Centre for Astrophysics \& Supercomputing, Swinburne University of Technology, P.O. Box 218, Hawthorn, VIC 3122, Australia\\
$^{2}$ Center for Astrophysics $|$ Harvard \& Smithsonian, 60 Garden Street, Cambridge, MA 02138, USA\\
$^{3}$ Department of Physics, The University of Texas at Dallas, 800 W. Campbell Rd., Richardson, TX 75080, USA\\
$^{4}$ Physics Dept., Boston University, 590 Commonwealth Avenue, Boston, MA 02215, USA\\
$^{5}$ Dipartimento di Fisica ``Aldo Pontremoli'', Universit\`a degli Studi di Milano, Via Celoria 16, I-20133 Milano, Italy\\
$^{6}$ Department of Physics \& Astronomy, University College London, Gower Street, London, WC1E 6BT, UK\\
$^{7}$ Lawrence Berkeley National Laboratory, 1 Cyclotron Road, Berkeley, CA 94720, USA\\
$^{8}$ Instituto de F\'{\i}sica, Universidad Nacional Aut\'{o}noma de M\'{e}xico,  Circuito de la Investigaci\'{o}n Cient\'{\i}fica, Ciudad Universitaria, Cd. de M\'{e}xico  C.~P.~04510,  M\'{e}xico\\
$^{9}$ NSF NOIRLab, 950 N. Cherry Ave., Tucson, AZ 85719, USA\\
$^{10}$ University of California, Berkeley, 110 Sproul Hall \#5800 Berkeley, CA 94720, USA\\
$^{11}$ Departamento de F\'isica, Universidad de los Andes, Cra. 1 No. 18A-10, Edificio Ip, CP 111711, Bogot\'a, Colombia\\
$^{12}$ Observatorio Astron\'omico, Universidad de los Andes, Cra. 1 No. 18A-10, Edificio H, CP 111711 Bogot\'a, Colombia\\
$^{13}$ Fermi National Accelerator Laboratory, PO Box 500, Batavia, IL 60510, USA\\
$^{14}$ Department of Astronomy and Astrophysics, UCO/Lick Observatory, University of California, 1156 High Street, Santa Cruz, CA 95064, USA\\
$^{15}$ Center for Cosmology and AstroParticle Physics, The Ohio State University, 191 West Woodruff Avenue, Columbus, OH 43210, USA\\
$^{16}$ Department of Physics, The Ohio State University, 191 West Woodruff Avenue, Columbus, OH 43210, USA\\
$^{17}$ The Ohio State University, Columbus, 43210 OH, USA\\
$^{18}$ School of Mathematics and Physics, University of Queensland, Brisbane, QLD 4072, Australia\\
$^{19}$ Centro de Investigaciones Energ\'eticas, Medioambientales y Tecnol\'ogicas (CIEMAT), Avenida Complutense 40, E-28040 Madrid, Spain \\
$^{20}$ Aix Marseille Univ, CNRS, CNES, LAM, Marseille, France\\
$^{21}$ Department of Physics, Southern Methodist University, 3215 Daniel Avenue, Dallas, TX 75275, USA\\
$^{22}$ Department of Physics and Astronomy, University of California, Irvine, 92697, USA\\
$^{23}$ Department of Physics and Astronomy, University of Waterloo, 200 University Ave W, Waterloo, ON N2L 3G1, Canada\\
$^{24}$ Perimeter Institute for Theoretical Physics, 31 Caroline St. North, Waterloo, ON N2L 2Y5, Canada\\
$^{25}$ Waterloo Centre for Astrophysics, University of Waterloo, 200 University Ave W, Waterloo, ON N2L 3G1, Canada\\
$^{26}$ Department of Physics, American University, 4400 Massachusetts Avenue NW, Washington, DC 20016, USA\\
$^{27}$ Department of Physics, University of Michigan, 450 Church Street, Ann Arbor, MI 48109, USA\\
$^{28}$ Department of Astronomy and Astrophysics, University of California, Santa Cruz, 1156 High Street, Santa Cruz, CA 95065, USA\\
$^{29}$ Departament de F\'{i}sica, Serra H\'{u}nter, Universitat Aut\`{o}noma de Barcelona, 08193 Bellaterra (Barcelona), Spain\\
$^{30}$ Institut de F\'{i}sica d’Altes Energies (IFAE), The Barcelona Institute of Science and Technology, Edifici Cn, Campus UAB, 08193, Bellaterra (Barcelona), Spain\\
$^{31}$ Instituci\'{o} Catalana de Recerca i Estudis Avan\c{c}ats, Passeig de Llu\'{\i}s Companys, 23, 08010 Barcelona, Spain\\
$^{32}$ Department of Physics and Astronomy, Siena College, 515 Loudon Road, Loudonville, NY 12211, USA\\
$^{33}$ Departamento de F\'{\i}sica, DCI-Campus Le\'{o}n, Universidad de Guanajuato, Loma del Bosque 103, Le\'{o}n, Guanajuato C.~P.~37150, M\'{e}xico.\\
$^{34}$ Instituto Avanzado de Cosmolog\'{\i}a A.~C., San Marcos 11 - Atenas 202. Magdalena Contreras. Ciudad de M\'{e}xico C.~P.~10720, M\'{e}xico\\
$^{35}$ Departament de F\'isica, EEBE, Universitat Polit\`ecnica de Catalunya, c/Eduard Maristany 10, 08930 Barcelona, Spain\\
$^{36}$ CIEMAT, Avenida Complutense 40, E-28040 Madrid, Spain\\
$^{37}$ Institute for Astronomy, University of Edinburgh, Royal Observatory, Blackford Hill, Edinburgh EH9 3HJ, UK\\
$^{38}$ Department of Physics and Astronomy, Sejong University, 209 Neungdong-ro, Gwangjin-gu, Seoul 05006, Republic of Korea\\
$^{39}$ Max Planck Institute for Extraterrestrial Physics, Gie\ss enbachstra\ss e 1, 85748 Garching, Germany\\
$^{40}$ Department of Astronomy, Tsinghua University, 30 Shuangqing Road, Haidian District, Beijing, China, 100190\\
$^{41}$ University of Michigan, 500 S. State Street, Ann Arbor, MI 48109, USA
\normalsize

\appendix

\section{Gaussian covariance of shear and galaxy angular correlations}
\label{sec:gaussian}

\subsection{Angular correlations between the shear and galaxy density fields}
\label{sec:angcorr}

The distortion of images due to gravitational lensing, at a 2D position on the sky $\vo$, can be described by a matrix depending on a scalar convergence $\kappa$ and a two-component shear $(\gamma_1, \gamma_2)$ \citep[for reviews, see][]{2001PhR...340..291B, 2015RPPh...78h6901K, 2018ARA&A..56..393M}.  In the weak lensing approximation these quantities are inter-related, and can be expressed in terms of a single scalar lensing potential.  The principal observable is the shear field, traced by galaxy ellipticities.  It is convenient to project the shear components at a point into a ``tangential'' and ``cross'' component relative to a direction which forms an angle $\phi$ with a fiducial $x$-axis,
\begin{equation}
\begin{split}
\gamma_t(\vo,\phi) &= - \gamma_1(\vo) \cos{2\phi} - \gamma_2(\vo) \sin{2\phi} , \\
\gamma_\times(\vo,\phi) &= \gamma_1(\vo) \sin{2\phi} - \gamma_2(\vo) \cos{2\phi} .
\end{split}
\end{equation}
It is also convenient to separate the convergence field into E-modes, $\kappa_E(\vo)$, and B-modes, $\kappa_B(\vo)$, where B-modes are not sourced by gravitational lensing, but may arise from noise or other effects.  In the weak lensing approximation the Fourier transforms of the shear and convergence components are related by,
\begin{equation}
\begin{split}
\tilde{\gamma}_1(\vl) &= \tilde{\kappa}_E(\vl) \cos{2\psi} - \tilde{\kappa}_B(\vl) \sin{2\psi} , \\
\tilde{\gamma}_2(\vl) &= \tilde{\kappa}_E(\vl) \sin{2\psi} + \tilde{\kappa}_B(\vl) \cos{2\psi} ,
\end{split}
\label{eq:gamma12}
\end{equation}
where $\psi$ is the direction of the wavevector $\vl$.

Using these definitions, the correlation functions of the shear field between two points with angular separation $\vt$ can be written in terms of the tangential and cross components relative to the separation vector between those points, and hence to the angular power spectra of the convergence fields.  After averaging over directions of $\vt$ we obtain,
\begin{equation}
\begin{split}
    \xi_+(\theta) &= \langle \gamma_t \gamma_t \rangle(\theta) + \langle \gamma_\times \gamma_\times \rangle(\theta) =  \intvl \, \left[ C_{\kappa\kappa}(\vl) + C_{BB}(\vl) \right] \, J_0(\ell \theta) , \\
    \xi_-(\theta) &= \langle \gamma_t \gamma_t \rangle(\theta) - \langle \gamma_\times \gamma_\times \rangle(\theta) =  \intvl \, \left[ C_{\kappa\kappa}(\vl) - C_{BB}(\vl) \right] \, J_4(\ell \theta) ,
\end{split}
\label{eq:xipm}
\end{equation}
where $C_{\kappa\kappa}(\vl) = \tilde{\kappa}_E(\vl) \, \tilde{\kappa}^*_E(\vl) \, \Omega_s$ and $C_{BB}(\vl) = \tilde{\kappa}_B(\vl) \, \tilde{\kappa}^*_B(\vl) \, \Omega_s$, where $\Omega_s$ is the angular survey area, $J_n$ indicates a Bessel function of the first kind, and we have used the notation $\kappa_E \rightarrow \kappa$ and $\kappa_B \rightarrow B$.

We may also observe the projected galaxy overdensity field at a 2D position, $\delta_g(\vo)$, and form a correlation function of the shear components with respect to this position.  This correlation is known as the average (tangential or cross) shear, and using the above definitions it may be expressed in terms of power spectra as,
\begin{equation}
\begin{split}
    \gamma_t(\theta) &= \into \, \delta_g(\vo) \, \gamma_t(\vo+\vt) = \intvl \, C_{g\kappa}(\vl) \, J_2(\ell \theta) , \\
    \gamma_\times(\theta) &= \into \, \delta_g(\vo) \, \gamma_\times(\vo+\vt) = \intvl \, C_{gB}(\vl) \, J_2(\ell \theta) ,
\end{split}
\end{equation}
where $C_{g\kappa}(\vl) = \tilde{\delta}_g(\vl) \, \tilde{\kappa}^*_E(\vl) \, \Omega_s$ and $C_{gB}(\vl) = \tilde{\delta}_g(\vl) \, \tilde{\kappa}^*_B(\vl) \, \Omega_s$.  Finally, the auto-correlation function of the galaxy overdensity field is,
\begin{equation}
    w(\theta) = \into \, \delta_g(\vo) \, \delta_g(\vo+\vt) = \intvl \, C_{gg}(\vl) \, J_0(\ell \theta) ,
\end{equation}
where $C_{gg}(\vl) = \tilde{\delta}_g(\vl) \, \tilde{\delta}^*_g(\vl) \, \Omega_s$.  Although we will ultimately form covariances with the projected correlation function $w_p(R)$, specifying the covariances with the angular correlation function $w(\theta)$ is a necessary first step.

\subsection{Relations for the angular power spectra}

The total convergence in a given sky direction for an ensemble of sources with probability distribution $p_s(\chi)$, as a function of comoving distance $\chi$, is given in the weak lensing approximation as \citep[see also,][]{2001MNRAS.321..439G, 2004PhRvD..70d3009H, 2010A&A...523A...1J, 2018MNRAS.478.4277S},
\begin{equation}
  \kappa_E(\vo) = \overline{\rho}_m \int_0^\infty d\chi \, \overline{\Sigma_c^{-1}}(\chi) \, \delta_m(\chi,\vo) ,
\end{equation}
where $\overline{\rho}_m$ is the mean cosmic matter density, $\delta_m$ is the matter overdensity and the average inverse critical surface mass density is
\begin{equation}
  \overline{\Sigma_c^{-1}}(\chi) = \int_\chi^\infty d\chi_s \, p_s(\chi_s) \, \Sigma_c^{-1}(\chi,\chi_s) ,
\end{equation}
in terms of the inverse critical surface mass density $\Sigma_c^{-1}$ in comoving density units at a lens plane depending on lens and source distances $\chi_l$ and $\chi_s$,
\begin{equation}
  \Sigma_c^{-1}(\chi_l,\chi_s) = \left\{
\begin{array}{ll}
  \frac{4\pi G}{c^2} \, \frac{(1+z_l) \, \chi_l \, (\chi_s - \chi_l)}{\chi_s} & \chi_s > \chi_l , \\ 0 & \chi_s < \chi_l . \\
\end{array}
\right.
\label{eq:invsigc}
\end{equation}
The projected overdensity of a galaxy distribution with probability distribution $p_l(\chi)$ in a given sky direction is,
\begin{equation}
  \delta_g(\vo) = \int_0^\infty d\chi \, p_l(\chi) \, \delta_{g,3D}(\chi,\vo)
\end{equation}

Taking the Fourier transform of these relations, and applying the Limber approximation, allows us to express the angular power spectra between the convergence and galaxy fields in terms of the source and lens distributions.  For the convergence power spectrum we obtain, considering in general the convergence traced by two source populations with different redshift distributions,
\begin{equation}
  C^{12}_{\kappa\kappa}(\ell) = \overline{\rho}_m^2 \int_0^\infty d\chi \, \frac{\overline{\Sigma_{c,1}^{-1}}(\chi) \, \overline{\Sigma_{c,2}^{-1}}(\chi)}{\chi^2} \, P_{mm} \left( \frac{\ell}{\chi} , \chi \right) ,
\end{equation}
in terms of the 3D matter power spectrum $P_{mm}(k,\chi)$ as a function of wavenumber $k$ and distance $\chi$.  Connecting with the usual kernel for weak lensing studies,
\begin{equation}
    q_\kappa(\chi) = \overline{\rho}_m \, \overline{\Sigma_c^{-1}}(\chi) = \frac{3 \Omega_m H_0^2}{2 c^2} \frac{\chi}{a(\chi)} \int_\chi^\infty d\chi_s \, p_s(\chi_s) \, \frac{(\chi_s-\chi)}{\chi_s}
\label{eq:lenskern}
\end{equation}
where $a(\chi)$ is the cosmic scale factor and using $\frac{4\pi G \overline{\rho}_m}{c^2} = \frac{3 \, \Omega_m \, H_0^2}{2 \, c^2}$.  For the cross-power spectrum between the convergence and projected galaxy density fields we find,
\begin{equation}
  C_{g\kappa}(\ell) = \overline{\rho}_m \int_0^\infty d\chi \, \frac{p_l(\chi) \, \overline{\Sigma_c^{-1}}(\chi)}{\chi^2} \, P_{gm} \left( \frac{\ell}{\chi} , \chi \right) ,
\label{eq:clgk}
\end{equation}
where $P_{gm}$ is the 3D galaxy-mass cross-power spectrum.  Finally, for the power spectrum of the projected galaxy density we obtain,
\begin{equation}
  C_{gg}(\ell) = \int_0^\infty d\chi \, \frac{p_l^2(\chi)}{\chi^2} \, P_{gg} \left( \frac{\ell}{\chi} , \chi \right) ,
\label{eq:clgg}
\end{equation}
where $P_{gg}$ is the 3D galaxy auto-power spectrum.  We note for completeness that for the correlations involving B-modes, $C_{BB} = C_{gB} = 0$.  The kernel for galaxy overdensity is sometimes written $q_g(\chi) = p_l(\chi)$

The contribution to the power spectra from noise in the observational fields, assuming the shear field is measured using a source sample with angular density $\bar{n}_s$ and shape noise per ellipticity component $\sigma_e$, is,
\begin{equation}
N_\kappa = N_B = \sigma_e^2 / \bar{n}_s .
\end{equation}
The noise power spectrum of the projected galaxy overdensity field is,
\begin{equation}
N_g = 1 / \bar{n}_l ,
\end{equation}
where $\bar{n}_l$ is the angular density of the lens sample.

It is useful to note how Eq.~\ref{eq:clgk} and Eq.~\ref{eq:clgg} simplify when a galaxy sample forms a uniform narrow redshift slice of thickness $L_\parallel$ around $\chi = \chi_{\rm eff}$.  In this case $p(\chi) = 1/L_\parallel$ and we have,
\begin{equation}
  C^{\rm Nar}_{gg}(\ell) = \frac{1}{\chi_{\rm eff}^2 \, L_\parallel} P_{gg}\left(\frac{\ell}{\chi_{\rm eff}}\right) ,
\label{eq:clggnar}
\end{equation}
\begin{equation}
  C^{\rm Nar}_{g\kappa}(\ell) = \overline{\rho}_m \, \frac{\overline{\Sigma_c^{-1}}(\chi_{\rm eff})}{\chi_{\rm eff}^2} \, P_{gm} \left( \frac{\ell}{\chi_{\rm eff}} \right) .
\label{eq:clgknar}
\end{equation}

\subsection{Covariances of the shear and galaxy angular correlations}

The covariances between the correlation functions introduced in Sec.~\ref{sec:angcorr} may be computed from the covariances between the angular power spectra of different fields (labelled $A,B,C,D$) and samples (labelled $i,j,k,l$) using the general relation \citep{2017MNRAS.470.2100K},
\begin{equation}
  {\rm Cov} \left[ C_{AB}^{ij} , C_{CD}^{kl} \right] = \left[ C_{AC}^{ik} + \delta^K_{ik} \, \delta^K_{AC} \, N_A^i \right] \left[ C_{BD}^{jl} + \delta^K_{jl} \, \delta^K_{BD} \,  N_B^j \right] + \left[ C_{AD}^{il} + \delta^K_{il} \,  \delta^K_{AD} \, N_A^i \right] \left[ C_{BC}^{jk} + \delta^K_{jk} \, \delta^K_{BC} \, N_B^j \right]
\label{eq:covgeneral}
\end{equation}
where $N$ indicates the noise auto-power spectrum of a field.  Just quoting covariances between the same samples ($k=i$, $l=j$), we find for the correlation functions introduced above \citep[also see,][]{2008A&A...477...43J},
\begin{equation}
  {\rm Cov} \left[ \xi_+(\theta) , \xi_+(\theta') \right] = \frac{2}{\Omega_s} \intl \left\{ \left[ C_{\kappa\kappa}(\ell) + N_\kappa \right]^2 + N_\kappa^2 \right\} \, J_0(\ell \theta) \, J_0(\ell \theta') ,
\end{equation}
\begin{equation}
  {\rm Cov} \left[ \xi_-(\theta) , \xi_-(\theta') \right] = \frac{2}{\Omega_s} \intl \left\{ \left[ C_{\kappa\kappa}(\ell) + N_\kappa \right]^2 + N_\kappa^2 \right\} \, J_4(\ell \theta) \, J_4(\ell \theta') ,
\end{equation}
\begin{equation}
  {\rm Cov} \left[ \gamma_t(\theta) , \gamma_t(\theta') \right] = \frac{1}{\Omega_s} \intl \left\{ C_{g\kappa}^2(\ell) + \left[ C_{\kappa\kappa}(\ell) + N_\kappa \right] \left[ C_{gg}(\ell) + N_g \right] \right\} J_2(\ell \theta) \, J_2(\ell \theta') ,
\end{equation}
\begin{equation}
  {\rm Cov} \left[ \gamma_\times(\theta) , \gamma_\times(\theta') \right] = \frac{1}{\Omega_s} \intl \, N_\kappa \, \left[ C_{gg}(\ell) + N_g \right] \, J_2(\ell \theta) \, J_2(\ell \theta') ,
\end{equation}
\begin{equation}
  {\rm Cov} \left[ w_{gg}(\theta) , w_{gg}(\theta') \right] = \frac{2}{\Omega_s} \intl \left[ C_{gg}(\ell) + N_g \right]^2 \, J_0(\ell \theta) \, J_0(\ell \theta') .
\end{equation}
The cross-covariances between different correlation functions are,
\begin{equation}
  {\rm Cov} \left[ \xi_+(\theta) , \xi_-(\theta') \right] = \frac{2}{\Omega_s} \intl \left\{ \left[ C_{\kappa\kappa}(\ell) + N_\kappa \right]^2 - N_\kappa^2 \right\} \, J_0(\ell \theta) \, J_4(\ell \theta') ,
\end{equation}
\begin{equation}
  {\rm Cov} \left[ \xi_+(\theta) , \gamma_t(\theta') \right] = \frac{2}{\Omega_s} \intl \, \left[ C_{\kappa\kappa}(\ell) + N_\kappa \right] \, C_{g\kappa}(\ell) \, J_0(\ell \theta) \, J_2(\ell \theta') ,
\end{equation}
\begin{equation}
  {\rm Cov} \left[ \xi_+(\theta) , w(\theta') \right] = \frac{2}{\Omega_s} \intl \, C_{g\kappa}^2(\ell) \, J_0(\ell \theta) \, J_0(\ell \theta') ,
\end{equation}
\begin{equation}
  {\rm Cov} \left[ \xi_-(\theta) , \gamma_t(\theta') \right] = \frac{2}{\Omega_s} \intl \, \left[ C_{\kappa\kappa}(\ell) + N_\kappa \right] \, C_{g\kappa}(\ell) \, J_0(\ell \theta) \, J_2(\ell \theta') ,
\end{equation}
\begin{equation}
  {\rm Cov} \left[ \xi_-(\theta) , w(\theta') \right] = \frac{2}{\Omega_s} \intl \, C_{g\kappa}^2(\ell) \, J_4(\ell \theta) \, J_0(\ell \theta') ,
\end{equation}
\begin{equation}
  {\rm Cov} \left[ \gamma_t(\theta) , w(\theta') \right] = \frac{2}{\Omega_s} \intl \, C_{g\kappa}(\ell) \, \left[ C_{gg}(\ell) + N_g \right] J_2(\ell \theta) \, J_0(\ell \theta') .
\label{eq:covgtwt}
\end{equation}

\section{Super-sample covariance}
\label{sec:supersample}

``Super-sample covariance'' (SSC) arises from the correlations between the large-scale structure modes contributing towards the observed statistics, and modes whose wavelength is larger than the survey size \citep[e.g.,][]{2013PhRvD..87l3504T, 2014PhRvD..90j3530L, 2018JCAP...06..015B}.  The contribution to the covariance of the convergence and galaxy density power spectra from super-sample modes is given by \citep{2017MNRAS.470.2100K},
\begin{equation}
\begin{split}
    {\rm Cov}_{\rm SSC} \left[ C^{ij}_{\kappa\kappa}(\ell_1) , C^{kl}_{\kappa\kappa}(\ell_2) \right] &= \int d\chi \frac{q^i_\kappa(\chi) \, q^j_\kappa(\chi) \, q^k_\kappa(\chi) \, q^l_\kappa(\chi)}{\chi^4} \, \frac{\partial P_m}{\partial \delta_b} \left( \frac{\ell_1}{\chi} , \chi \right) \, \frac{\partial P_m}{\partial \delta_b} \left( \frac{\ell_2}{\chi} , \chi \right) \, \sigma_b(\chi) \\
    {\rm Cov}_{\rm SSC} \left[ C^{ij}_{g\kappa}(\ell_1) , C^{ik}_{g\kappa}(\ell_2) \right] &= \int d\chi \frac{[q^i_g(\chi)]^2 \, q^j_\kappa(\chi) \, q^k_\kappa(\chi)}{\chi^4} \left[ \frac{\partial P_m}{\partial \delta_b} \left( \frac{\ell_1}{\chi} , \chi \right) - b P_m \left( \frac{\ell_1}{\chi} , \chi \right) \right] \left[ \frac{\partial P_m}{\partial \delta_b} \left( \frac{\ell_2}{\chi} , \chi \right) - b P_m \left( \frac{\ell_2}{\chi} , \chi \right) \right] \sigma_b(\chi) \\
    {\rm Cov}_{\rm SSC} \left[ C_{gg}(\ell_1) , C_{gg}(\ell_2) \right] &= \int d\chi \frac{q^4_g(\chi)}{\chi^4} \, \left[ \frac{\partial P_m}{\partial \delta_b} \left( \frac{\ell_1}{\chi} , \chi \right) - 2 b P_m \left( \frac{\ell_1}{\chi} , \chi \right) \right] \left[ \frac{\partial P_m}{\partial \delta_b} \left( \frac{\ell_2}{\chi} , \chi \right) - 2 b P_m \left( \frac{\ell_2}{\chi} , \chi \right) \right] \sigma_b(\chi) ,
\end{split}
\label{eq:clcovssc}
\end{equation}
where $q_\kappa(\chi) = \overline{\rho}_m \, \overline{\Sigma_c^{-1}}(\chi)$ is the lensing kernel given by Eq.~\ref{eq:lenskern}, $q_g(\chi) = p_l(\chi)$ is the galaxy overdensity kernel, $\partial P_m/\partial \delta_b$ is the ``response function'' of the matter power spectrum to the background mode, $b$ is the linear galaxy bias, and $\sigma_b(\chi)$ is the variance of the background mode over the survey window.  The cross-covariances between different angular power spectra follow by combining their corresponding contributions to the integrands of Eq.~\ref{eq:clcovssc}.  These terms are evaluated by, for the variance of the background mode,
\begin{equation}
    \sigma_b(\chi) = \intkperp \, P_{\rm lin}(k_\perp,\chi) \, |\tilde{W}_s(k_\perp\chi)|^2 ,
\end{equation}
in terms of the linear matter power spectrum $P_{\rm lin}(k)$ at comoving distance $\chi$, and the angular power spectrum of the survey window projected at the co-moving distance,
\begin{equation}
    \tilde{W}_s(\ell) = \int \frac{d\psi}{2\pi} \into \, W(\vo) \, e^{i \vl \cdot \vo} ,
\end{equation}
where $W(\vo)$ is the survey window function and $\psi$ is the direction of wavevector $\vl$.  For a disk-like geometry, $|\tilde{W}_s(\ell)|^2 = \left[ 2 J_1(\ell \theta_s) / (\ell \theta_s) \right]^2$ where $\pi \theta_s^2 = \Omega_s$ (we use the window function of each Buzzard region in the calculation for this paper).  The response function of the matter power spectrum is given in terms of a galaxy halo model description by,
\begin{equation}
    \frac{\partial P_m}{\partial \delta_b}(k,\chi) = \left[ \frac{68}{21} - \frac{1}{3} \frac{d\ln{[k^3 P_{\rm lin}(k,\chi)]}}{d\ln{k}} \right] P_{\rm lin}(k,\chi) \left[ I_1^1(k) \right]^2 + I_2^1(k) ,
\end{equation}
in terms of the halo model factors, evaluated by integrating over halo mass $M$,
\begin{equation}
\begin{split}
I_1^1(k) &= \int dM \, \frac{dn}{dM} \, \frac{M}{\overline{\rho}_m} \, b(M) \, \tilde{u}_M(k) , \\
I_2^1(k) &= \int dM \, \frac{dn}{dM} \, \left( \frac{M}{\overline{\rho}_m} \right)^2 \, b(M) \, \left[ \tilde{u}_M(k) \right]^2 ,
\end{split}
\end{equation}
where $dn/dM$ is the halo mass function, $b(M)$ is the halo bias function, and $\tilde{u}_M(k)$ is the Fourier transform of the NFW halo mass profile.  We use the fitting formulae of \cite{2010ApJ...724..878T} for the halo mass and bias functions, the concentration-mass relation of \cite{2008MNRAS.390L..64D}, and extrapolate the integrals to zero mass following the method of \cite{2021A&A...646A.129J}.  The contribution of super-sample variance to the cosmic shear correlation function covariance is then deduced using Eq.~\ref{eq:xipm}, taking a single source sample as an example,
\begin{equation}
\begin{split}
  {\rm Cov}_{\rm SSC} \left[ \xi^{ii}_+(\theta) , \xi^{ii}_+(\theta') \right] &= \intl \intlp \, J_0(\ell\theta) \, J_0(\ell'\theta') \, {\rm Cov}_{\rm SSC} \left[ C^{ii}_{\kappa\kappa}(\ell) , C^{ii}_{\kappa\kappa}(\ell') \right] \\
  &= \int d\chi \, \frac{[q^i_\kappa(\chi)]^4}{\chi^4} \, \sigma_b(\chi) \, \overline{R}_0(\chi,\theta) \, \overline{R}_0(\chi,\theta') ,
\end{split}
\end{equation}
where we introduced the Bessel-function average of the response function,
\begin{equation}
  \overline{R}_n(\chi,\theta) = \intl \, J_n(\ell\theta) \, \frac{\partial P_m}{\partial \delta_b} \left( \frac{\ell}{\chi} , \chi \right) .
\end{equation}
Likewise,
\begin{equation}
    {\rm Cov}_{\rm SSC} \left[ \xi^{ii}_-(\theta) , \xi^{ii}_-(\theta') \right] = \int d\chi \, \frac{[q^i_\kappa(\chi)]^4}{\chi^4} \, \sigma_b(\chi) \, \overline{R}_4(\chi,\theta) \, \overline{R}_4(\chi,\theta') .
\end{equation}
The super-sample contribution to the covariance of the average tangential shear between lens sample $i$ and source sample $j$ is,
\begin{equation}
\begin{split}
  {\rm Cov}_{\rm SSC} \left[ \gamma^{ij}_t(\theta) , \gamma^{ij}_t(\theta') \right] &= \intl \intlp \, J_2(\ell\theta) \, J_2(\ell'\theta') \, {\rm Cov}_{\rm SSC} \left[ C^{ij}_{g\kappa}(\ell) , C^{ij}_{g\kappa}(\ell') \right] \\
  &= \int d\chi \, \frac{[q^i_g(\chi)]^2 \, [q^j_\kappa(\chi)]^2}{\chi^4} \, \sigma_b(\chi) \left[ \overline{R}_2(\chi,\theta) - b \, \overline{P}_2(\chi,\theta) \right] \left[ \overline{R}_2(\chi,\theta') - b \, \overline{P}_2(\chi,\theta') \right] ,
\end{split}
\end{equation}
where we introduced the Bessel-function average of the matter power spectrum,
\begin{equation}
  \overline{P}_n(\chi,\theta) = \intl \, J_n(\ell\theta) \, P_m \left( \frac{\ell}{\chi} , \chi \right) .
\end{equation}
Finally, the super-sample contribution to the covariance of the angular correlation function is,
\begin{equation}
\begin{split}
  {\rm Cov}_{\rm SSC} \left[ w(\theta) , w(\theta') \right] &= \intl \intlp \, J_0(\ell\theta) \, J_0(\ell'\theta') \, {\rm Cov}_{\rm SSC} \left[ C_{gg}(\ell) , C_{gg}(\ell') \right] \\
  &= \int d\chi \, \frac{q^4_g(\chi)}{\chi^4} \, \sigma_b(\chi) \left[ \overline{R}_0(\chi,\theta) - 2 \, b \, \overline{P}_0(\chi,\theta) \right] \left[ \overline{R}_0(\chi,\theta') - 2 \, b \, \overline{P}_0(\chi,\theta') \right] .
\end{split}
\end{equation}
The SSC contribution to the cross-covariances between different correlation functions follow by generalising these expressions in a natural way to combine their respective contributions to the integrands.

\section{Noise correction}
\label{sec:noisecorr}

The analytical estimate of the Gaussian noise contribution to the covariance is approximate, because it neglects the effects of survey boundaries and varying completeness on the measured number of pairs of objects \citep{2002A&A...396....1S,2018MNRAS.479.4998T,2021A&A...646A.129J}.  In our covariance estimates we replace these analytical noise forecasts with direct pair counts of weighted objects in each separation bin.  For the cosmic shear correlation functions, the direct noise estimate is given by,
\begin{equation}
   {\rm Cov}_{\rm Noise} \left[ \xi^{ij}_+(\theta) , \xi^{ij}_+(\theta) \right] = {\rm Cov}_{\rm Noise} \left[ \xi^{ij}_-(\theta) , \xi^{ij}_-(\theta) \right] = \frac{2 \, (1 + \delta^K_{ij}) \, (\sigma^i_e)^2 \, (\sigma^j_e)^2}{N_{\rm pair,eff}(\theta)} ,
\end{equation}
where $\sigma^i_e$ is the shape noise of shear sample $i$, and the effective source pair count in the separation bin is given by,
\begin{equation}
    N_{\rm pair,eff}(\theta) = \frac{\left( \sum_{ij,{\rm bin}} w_i \, w_j \right)^2}{\sum_{ij,{\rm bin}} w_i^2 \, w_j^2} ,
\label{eq:effpair}
\end{equation}
where the sum is over all pairs of sources in the separation bin, where the sources have weights $w_i$.  For the noise contribution to the average tangential shear between lens sample $i$ and source sample $j$ we use,
\begin{equation}
    {\rm Cov}_{\rm Noise} \left[ \gamma^{ij}_t(\theta) , \gamma^{ij}_t(\theta) \right] = \frac{(\sigma^j_e)^2}{N_{\rm pair,eff}(\theta)} ,
\end{equation}
where the effective pair count has the same form as Eq.~\ref{eq:effpair}, now using the respective weights of the lens and source sample.  The noise contribution to the angular galaxy clustering is,
\begin{equation}
    {\rm Cov}_{\rm Noise} \left[ w(\theta) , w(\theta) \right] = \frac{2}{N_{\rm pair,eff}(\theta)} ,
\end{equation}
where the effective pair count is now evaluated for all pairs of the lens sample using Eq.~\ref{eq:effpair}.  For the projected correlation function we use,
\begin{equation}
    {\rm Cov}_{\rm Noise} \left[ w_p(R) , w_p(R) \right] = \frac{8 \, \Pi^2_{\rm max}}{N_{\rm pair,eff}(R)} ,
\end{equation}
where $\Pi_{\rm max}$ is the largest line-of-sight separation used in the measurement.  In each case, the analytical covariance for correlation function $\xi$ is then corrected as,
\begin{equation}
    {\rm Cov} \left[ \xi , \xi \right] \rightarrow {\rm Cov} \left[ \xi , \xi \right] - {\rm Cov}_{\rm Noise,Ana} \left[ \xi , \xi \right] + {\rm Cov}_{\rm Noise} \left[ \xi , \xi \right] ,
\end{equation}
where the terms ${\rm Cov}_{\rm Noise,Ana} \left[ \xi , \xi \right]$ are evaluated by setting the power spectra to zero in the expressions for the analytical covariances.

\section{Covariances involving the projected correlation function}
\label{sec:covwp}

The projected galaxy correlation function at projected separation $\vR$ is defined by,
\begin{equation}
  w_p(\vR) = \int_{-\Pi_{\rm max}}^{\Pi_{\rm max}} d\Pi \, \xi_{gg} (\vR,\Pi) ,
\end{equation}
where $\Pi_{\rm max}$ is the upper limit used in the measurement, and $\xi_{gg}(\vs)$ is the 3D galaxy correlation function.  We can express this in terms of the power spectrum by using $\xi_{gg}(\vs) = \intk \, P_{gg}(\vk) \, e^{-i\vk \cdot \vs}$, obtaining,
\begin{equation}
  w_p(\vR) = \intk P_{gg}(\vk) \, e^{-i\vk_\perp \cdot \vR} \int_{-\Pi_{\rm max}}^{\Pi_{\rm max}} d\Pi\, e^{-i k_\parallel \Pi} .
\end{equation}
Averaging the measurement over directions $\phi$ we find,
\begin{equation}
  w_p(R) = 2 \Pi_{\rm max} \intk \, P_{gg}(\vk) \, J_0(k_\perp R) \, j_0(k_\parallel\Pi_{\rm max}) .
\label{eq:wppk}
\end{equation}
In the Limber approximation we suppose that only tangential modes contribute to the projected clustering, which allows us to simplify Eq.~\ref{eq:wppk} to the form,
\begin{equation}
\begin{split}
  w_p^{\rm Lim}(R) &= \intvkperp P_{gg}(\vk_\perp) \, J_0(k_\perp R) \times 2 \Pi_{\rm max} \, \intkpar \, j_0(k_\parallel\Pi_{\rm max}) \\
&= \intvkperp \, P_{gg}(\vk_\perp) \, J_0(k_\perp R) .
\end{split}
\label{eq:wplimpk}
\end{equation}

The covariance of $w_p(R)$ follows from Eq.~\ref{eq:wppk},
\begin{equation}
  {\rm Cov} \left[ w_p(R), w_p(R') \right] = 4 \Pi_{\rm max}^2 \intk \intkp {\rm Cov} \left[ P_{gg}(\vk), P_{gg}(\vk') \right] J_0(k_\perp R) \, J_0(k'_\perp R') \, j_0(k_\parallel \Pi_{\rm max}) \, j_0(k'_\parallel\Pi_{\rm max}) .
\end{equation}
Using ${\rm Cov} \left[ P_{gg}(\vk), P_{gg}(\vk') \right] = 2 \, \left[ P_{gg}(\vk) + \frac{1}{n_g} \right]^2 \, \tilde{\delta}_D(\vk-\vk')$, where the factor of 2 accounts for the fact that the modes $-\vk$ and $\vk$ are not independent, this becomes,
\begin{equation}
  {\rm Cov} \left[ w_p(R), w_p(R') \right] = \frac{8 \Pi_{\rm max}^2}{V_s} \intk \left[ P_{gg}(\vk) + \frac{1}{n_g} \right]^2 \, J_0(k_\perp R) \, J_0(k_\perp R') \, j_0^2(k_\parallel\Pi_{\rm max}) ,
\end{equation}
where $V_s$ is the survey volume.  In the Limber approximation where only tangential modes contribute to the covariance,
\begin{equation}
\begin{split}
  {\rm Cov}^{\rm Lim} \left[ w_p(R), w_p(R') \right] &= \frac{8 \Pi_{\rm max}^2}{V_s} \intvkperp \left[ P_{gg}(k_\perp) + \frac{1}{n_g} \right]^2 \, J_0(k_\perp R) \, J_0(k_\perp R') \, \intkpar j_0^2(k_\parallel\Pi_{\rm max}) \\
  &= \frac{4 \Pi_{\rm max}}{V_s} \intkperp \left[ P_{gg}(k_\perp) + \frac{1}{n_g} \right]^2 \, J_0(k_\perp R) \, J_0(k_\perp R') \\
  &= \frac{4 \Pi_{\rm max}}{V_s} \intkperp \left[ P_{gg}^2(k_\perp) + \frac{2 P_{gg}(k_\perp)}{n_g} \right] \, J_0(k_\perp R) \, J_0(k_\perp R') + \frac{2 \Pi_{\rm max}}{\pi R \, n_g^2 \, V_s} \, \delta_D(R-R') .
\label{eq:wpcov}
\end{split}
\end{equation}
We use the Limber-approximated covariance in Eq.\ref{eq:wpcov} as our fiducial evaluation.

The cross-covariance between $\gamma_t(\theta)$ and $w_p(R)$ may be deduced by converting $w_p(R)$ to angular separations,
\begin{equation}
  {\rm Cov} \left[ \gamma_t(\theta) , w^{\rm Lim}_p(R') \right] = L_\parallel \, {\rm Cov} \left[ \gamma_t(\theta) , w(R'/\chi_{\rm eff}) \right] ,
\end{equation}
where the expression on the right-hand side is evaluated using Eq.~\ref{eq:covgtwt}.  The cross-covariance of the cosmic shear correlation functions $\xi_\pm(\theta)$ and galaxy projected correlation function $w_p(R)$ may be written using Eq.~\ref{eq:xipm} and Eq.~\ref{eq:wppk} as,
\begin{equation}
  {\rm Cov} \left[ \xi_\pm(\theta) , w_p(R) \right] = 2 \Pi_{\rm max} \intvl \intk \, {\rm Cov} \left[ C_{\kappa\kappa}(\vl) , P_{gg}(\vk) \right] \, J_{0/4}(\ell\theta) \, J_0(k_\perp R) \, j_0(k_\parallel\Pi_{\rm max}) .
\end{equation}
Now we consider,
\begin{equation}
\begin{split}
  {\rm Cov} \left[ C_{\kappa\kappa}(\vl) , P_{gg}(\vk) \right] &= \overline{\rho}_m^2 \int_0^\infty d\chi \, \frac{\overline{\Sigma_{c,1}^{-1}}(\chi) \, \overline{\Sigma_{c,2}^{-1}}(\chi)}{\chi^2} \, {\rm Cov} \left[ P_{mm} \left( \frac{\vl}{\chi} , \chi \right) , P_{gg}(\vk) \right] \\
  &= \overline{\rho}_m^2 \, L_\parallel \, \frac{\overline{\Sigma_{c,1}^{-1}}(\chi_{\rm eff}) \, \overline{\Sigma_{c,2}^{-1}}(\chi_{\rm eff}) }{\chi_{\rm eff}^2} \, {\rm Cov} \left[ P_{mm}(\vk'_\perp) , P_{gg}(\vk_\perp) \right] \, \tilde{\delta}_D(k_\parallel) \\
  &= \overline{\rho}_m^2 \, L_\parallel \, \frac{ \overline{\Sigma_{c,1}^{-1}}(\chi_{\rm eff}) \, \overline{\Sigma_{c,2}^{-1}}(\chi_{\rm eff})}{\chi_{\rm eff}^2} \, 2 \, P^2_{gm}(\vk_\perp) \, \tilde{\delta}_D(k_\parallel) \, \tilde{\delta}_D \left( \frac{\vl}{\chi_{\rm eff}} - \vk_\perp \right) .
\end{split}
\label{eq:covclkkpgg}
\end{equation}
where the covariance is non-zero over the range of the narrow redshift slice of galaxies, where $\vl = \vk_\perp \, \chi_{\rm eff}$.  Hence,
\begin{equation}
    {\rm Cov} \left[ \xi_\pm(\theta) , w_p(R) \right] = 2 \, \overline{\rho}_m^2 \, \frac{ \overline{\Sigma_{c,1}^{-1}}(\chi_{\rm eff}) \, \overline{\Sigma_{c,2}^{-1}}(\chi_{\rm eff})}{\chi_{\rm eff}^2} \intkperp P^2_{gm}(k_\perp) \, J_{0/4}(k_\perp \chi_{\rm eff} \theta) \, J_0(k_\perp R) .
\end{equation}

\section{Code comparisons}
\label{sec:codecomparison}

\begin{figure*}
\centering
\includegraphics[width=0.9\columnwidth]{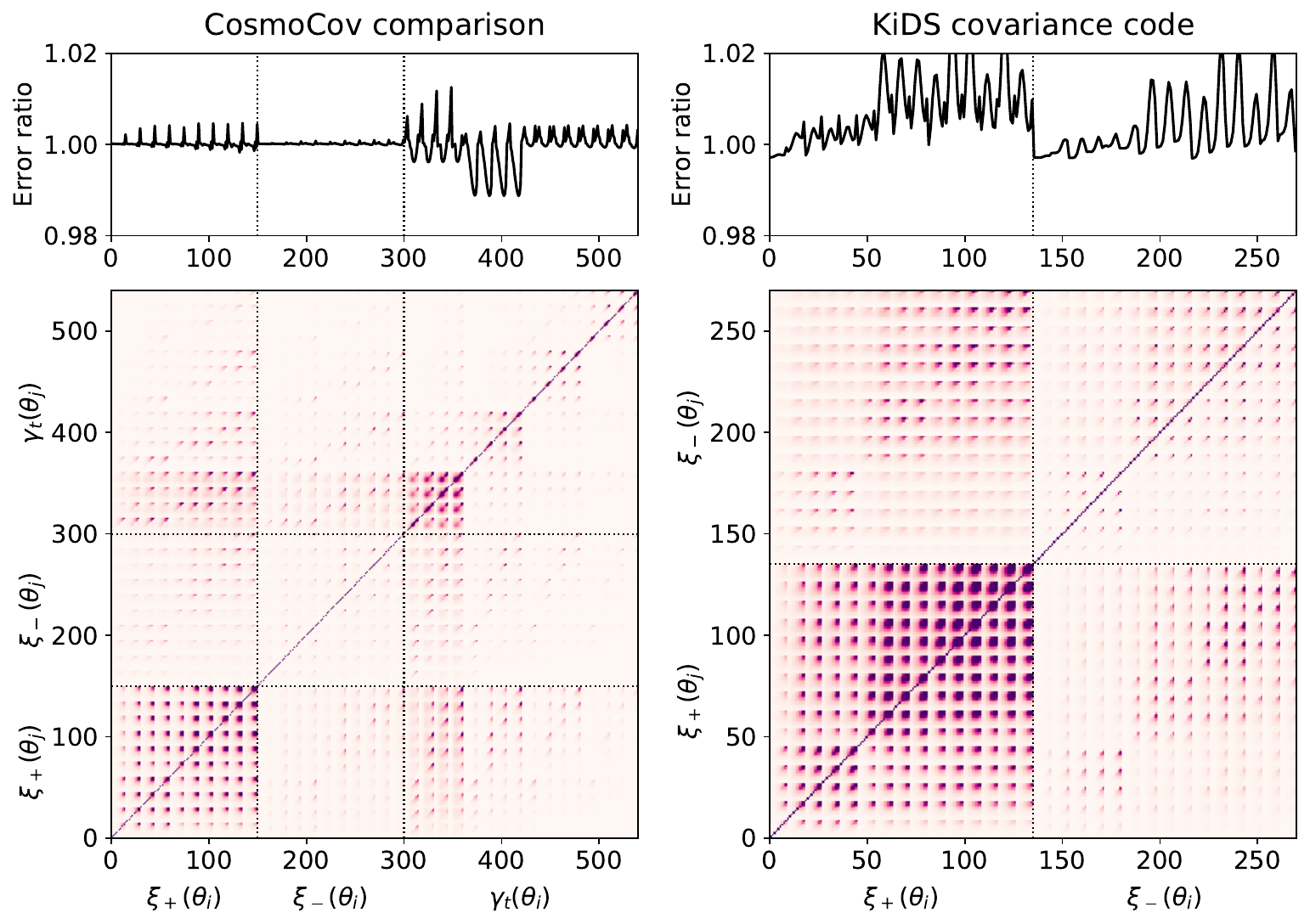}
\caption{The results of a code comparison between our analytical covariance and the evaluations of two external codes: {\tt CosmoCov} (left-hand panels) and the KiDS covariance code (right-hand panels).  The comparison with {\tt CosmoCov} includes the $\xi_{\pm}(\theta)$ and $\gamma_t(\theta)$ correlations, and the comparison with the KiDS code includes the $\xi_{\pm}(\theta)$ correlations.  The correlation functions across source (and lens) bins are arranged into a data vector of length (540, 270) for the ({\tt CosmoCov}, KiDS) configurations.  The upper panels show the ratio of diagonal errors between our code and the external code, and the lower panels display a comparison of the full covariance matrices, where our evaluation is depicted in the upper-left triangle, and the evaluation of the external code is shown in the lower-right triangle.}
\label{fig:covcomparison}
\end{figure*}

We compared our analytical covariance evaluations to those of two existing codes: {\tt CosmoCov} \citep{2017MNRAS.470.2100K,2020MNRAS.497.2699F}, and the code used in the KiDS-1000 cosmology analysis \citep{2021A&A...646A.129J}.  For the comparison with {\tt CosmoCov} we considered the $\xi_{\pm}(\theta)$ and $\gamma_t(\theta)$ correlations, using a test configuration corresponding to the DES-Y3 source samples ($N_{\rm tom} = 4$) and four DESI spectroscopic redshift bins ($N_{\rm lens} = 4$) with divisions $z = (0.1, 0.3, 0.5, 0.7, 0.9)$, using $N_{\rm sep} = 15$ logarithmically-spaced angular separation bins in the range $0.003 < \theta < 3 \, {\rm deg}$.  The total data vector in this case has length $N_{\rm tom} \left( N_{\rm tom} + 1 \right) N_{\rm sep} + N_{\rm lens} N_{\rm tom} N_{\rm sep} = 540$.  For the comparison with the KiDS covariance code we considered $\xi_{\pm}(\theta)$ correlations for the configuration of KiDS-1000 shear correlation function measurements, using $N_{\rm tom} = 5$ and $N_{\rm sep} = 9$ logarithmically-spaced angular separation bins in the range $0.5 < \theta < 300 \, {\rm arcmin}$.  The total data vector in this case has length $N_{\rm tom} \left( N_{\rm tom} + 1 \right) N_{\rm sep} = 270$.  The covariance evaluations include the Gaussian, super-sample and noise contributions in all cases.  Fig.~\ref{fig:covcomparison} displays the results of both code comparisons.  The upper panels show the ratio of diagonal errors between our code and the external code, where we find agreement to better than $1\%$ accuracy.  The residual differences are due to slightly different implementations of the halo model combined with numerical factors such as integration limits and interpolation methods.  The lower panels display a comparison of the full covariance matrices, where our evaluation is depicted in the upper-left triangle, and the evaluation of the external code is shown in the lower-right triangle.  The covariance matrices agree closely in all respects.

\section{Best-fit correlation functions}
\label{sec:fixed_cosmo_fits}

\begin{figure*}
\centering
\includegraphics[width=1.0\columnwidth]{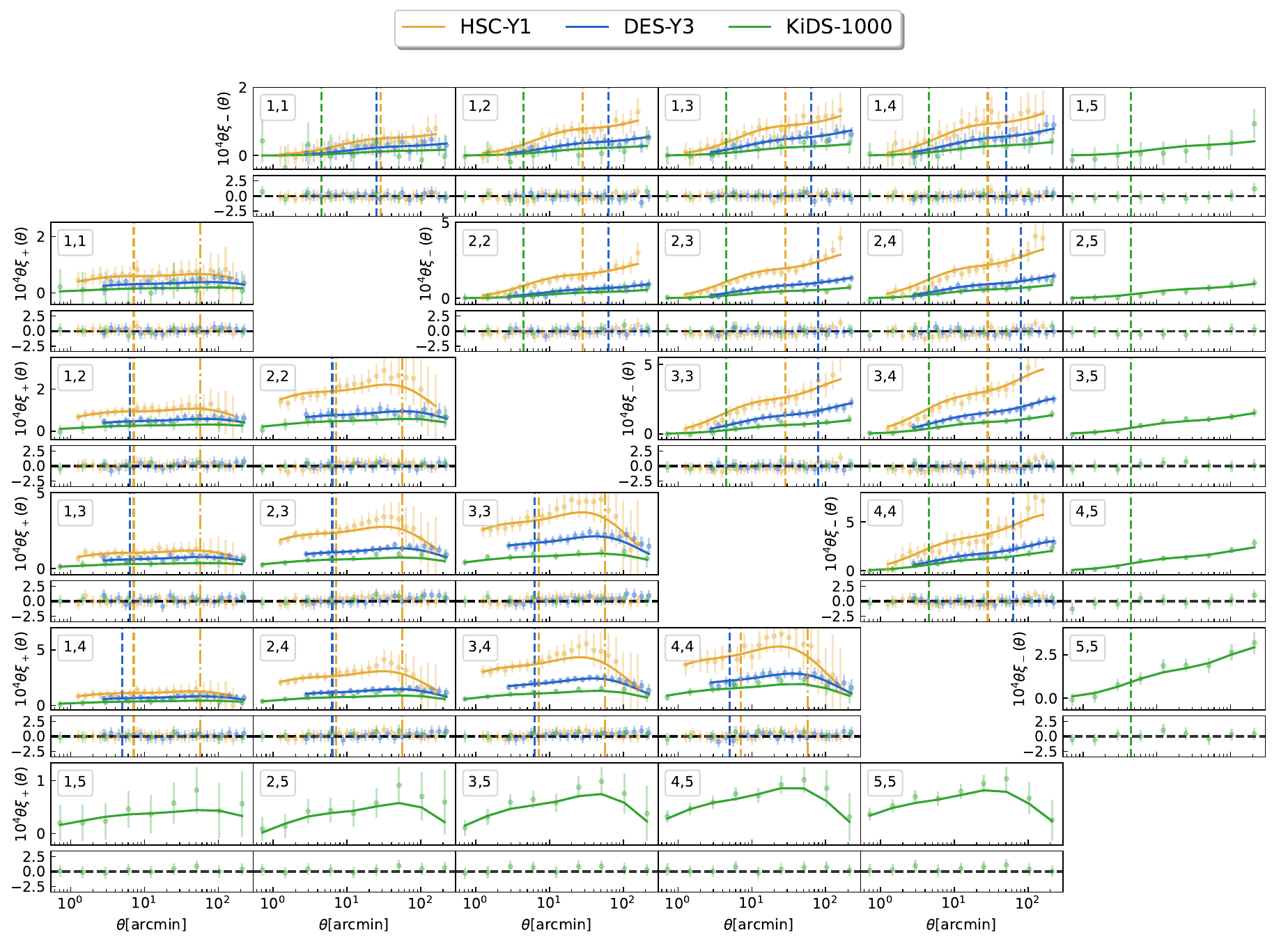}
\caption{The mock mean $\xi_\pm$ measurements and best-fitting models for the $3 \times 2$-pt fits. Each panel shows the measured correlation function for a combination of tomographic source bins averaged over 8 Buzzard mocks including overlapping and non-overlapping regions corresponding to cosmic shear, GGL and clustering.  The errors are determined by the analytical covariance and the solid curves represent the best-fitting model. The dashed vertical lines represent the lower scale cuts and the dashed-dotted vertical lines correspond to upper scale cuts (only for HSC-Y1). The smaller panels represent the fractional deviation between the predicted and measured correlation functions within each bin.  The HSC-Y1 data is represented in yellow, whilst the results for DES-Y3 and KiDS-1000 are shown in blue and green, respectively.}
\label{fig:xipm_best_fit}
\vspace{0.5cm}
\end{figure*}

\begin{figure*}
\centering
\includegraphics[width=1.0\columnwidth]{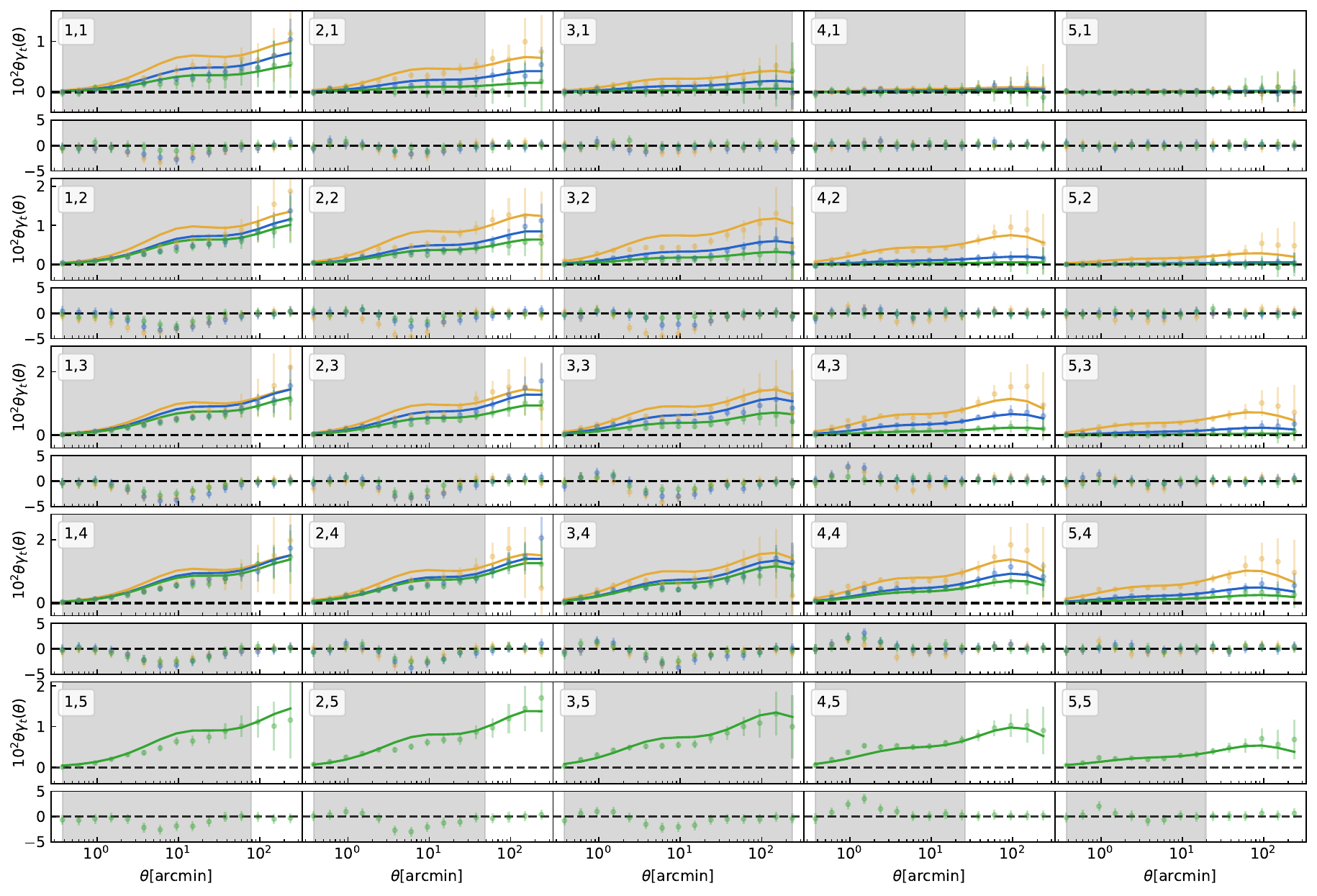}
\caption{The mock mean $\gamma_t$ measurements and best-fitting models for the $3 \times 2$-pt fits, presented in the same style as Fig.\ref{fig:xipm_best_fit}.  Each panel shows the results for a combination of source bin (rows) and lens bin (columns).  The same scale cuts are used for the three lensing surveys, and we represent them as grey regions covering excluded data.}
\label{fig:gt_best_fit}
\vspace{0.5cm}
\end{figure*}

\begin{figure*}
\centering
\includegraphics[width=0.5\columnwidth]{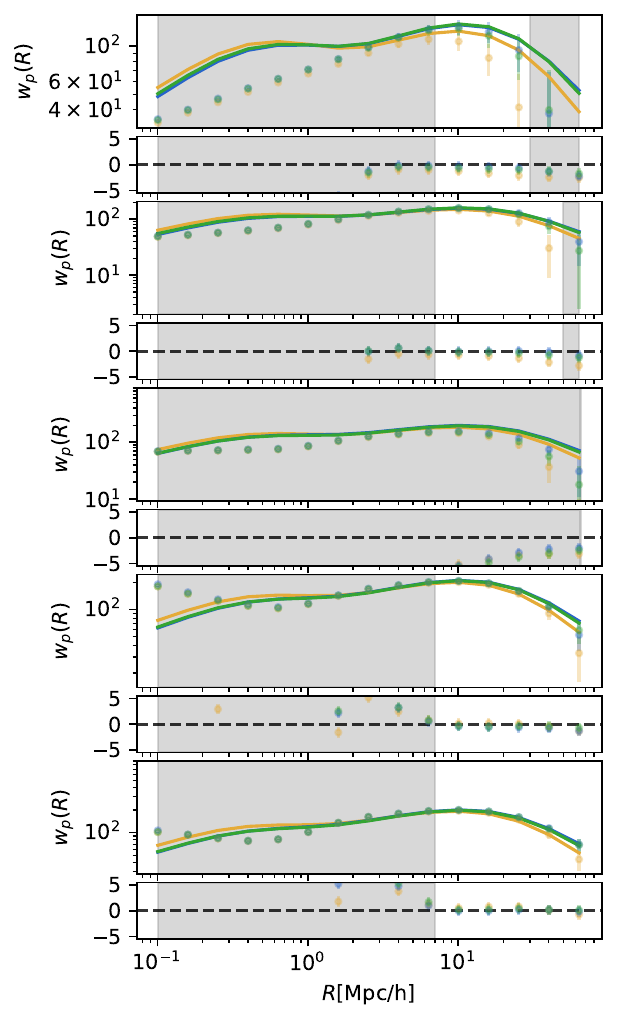}
\caption{The mock mean $w_p$ measurements and best-fitting models for the $3 \times 2$-pt fits, presented in the same style as Fig.\ref{fig:xipm_best_fit}.  The separate panels represent the five different DESI lens bins from low redshift (top) to high redshift (bottom). The excluded separations are represented as the grey region, assuming a $7 \, h^{-1}$ Mpc scale cut.  We note that the third lens redshift bin is known to be affected by the light cone transition used in the Buzzard simulations, and is excluded from our fits.}
\label{fig:wp_best_fit}
\vspace{0.5cm}
\end{figure*}

The best-fitting $3\times 2$-pt correlation functions for the tests described in Sec.~\ref{sec:fixedcosmo} are displayed in Fig.~\ref{fig:xipm_best_fit} (cosmic shear), Fig.~\ref{fig:gt_best_fit} (galaxy-galaxy lensing), and Fig.~\ref{fig:wp_best_fit} (galaxy clustering).

\end{document}